\documentclass[graphicx,preprint,11pt]{aastex}
\input{psfig}
\ifx\pdfoutput\@undefined\usepackage[usenames,dvips]{color}
\else\usepackage[usenames,dvipsnames]{color}
\IfFileExists{pdfcolmk.sty}{\usepackage{pdfcolmk}}{} 
\fi

\lefthead{}
\righthead{}

\begin{document}

\title{Dynamical evolution of planetary systems}

\author{\textbf{Alessandro Morbidelli}}
\affil{\small\em Dep. Cassiopee, University of Nice - Sophia Antipolis, CNRS, Observatoire de la C\^{o}te d'Azur; Nice, France}

\newpage

\begin{abstract}

The apparent regularity of the motion of the giant planets of our
solar system suggested for decades that said planets formed onto
orbits similar to the current ones and that nothing dramatic ever
happened during their lifetime. The discovery of extra-solar planets
showed astonishingly that the orbital structure of our planetary
system is not typical.  Many giant extra-solar planets have orbits
with semi major axes of $\sim 1$~AU, and some have even smaller
orbital radii, sometimes with orbital periods of just a few
days. Moreover, most extra-solar planets have large eccentricities, up
to values that only comets have in our solar system. Why such a big
diversity between our solar system and the extra-solar systems, as
well as among the extra-solar systems themselves?  This chapter aims
to give a partial answer to this fundamental question. It's guideline
is a discussion of the evolution of our solar system, certainly biased
by a view that emerges, in part, from a series of works in which I
have been involved. According to this view, the giant planets of the
solar system suffered radial migration while they were still embedded in a
proto-planetary disk of gas and presumably achieved a multi-resonant
orbital configuration, characterized by smaller inter-orbital spacings
and smaller eccentricities and inclinations with respect to the
current configuration. The current orbits of the giant planets have
been achieved during a phase of orbital instability, during which the
planets acquired temporarily large-eccentricity orbits and all
experienced close encounters with at least another planet. This
instability phase occurred presumably during the putative ``Late Heavy
Bombardment'' of the terrestrial planets, approximately $\sim 3.9$~Gy
ago (Tera et al., 1974). The interaction with a massive distant
planetesimal disk (the ancestor of the current Kuiper belt) eventually
damped the eccentricities of the planets, ending the phase of mutual
planetary encounters and parking the planets onto their current,
stable orbits. This new view of the evolution of the solar system,
makes our system not very different from the extra-solar ones. In
fact, the best explanation for the large orbital eccentricities of
extra-solar planets is that the planets that are observed are the
survivors of strong instability phases of original multi-planet
systems on quasi-circular orbits. The main difference between the
solar system and the extra-solar systems is in the magnitude of such
an instability. In the extra-solar systems, encounters among giant
planets had to be the norm. In our case, the two major planets
(Jupiter and Saturn) never had close encounters with each other: they
only encountered ``minor'' planets like Uranus and/or Neptune. This
was probably just mere luck, as simulations show that Jupiter-Saturn
encounters in principle could have occurred. Another relevant
difference with the extra-solar planets is that, during the gas-disk
phase, our planets avoided to migrate permanently into the inner solar
system, thanks to the specific mass ratio of the Jupiter/Saturn pair
and the rapid disappearance of the disk soon after the formation of
the giant planets. This chapter ends on a note on terrestrial
planets. The structure of a terrestrial-planet system depends
sensitively on the dynamical evolution of the giant planets and on
their final orbits. It appears clear that habitable terrestrial
planets, with moderate eccentricity orbits, cannot exist in systems
where the giant planets became violently unstable and developed very
elliptic orbits. Thus, our very existence is possible only because the
instability phase experienced by the giant planets of our solar system
was of ``moderate'' strength.

\end{abstract}

\section{Introduction}
\label{intro}

It is now clear that the planetary systems that we observe did not
form in their current configuration, but they have been heavily
modified by a non-trivial dynamical evolution. Processes like planet
migration, resonant trapping, planet-planet scattering, mutual
collisions, hyperbolic ejections have sculpted the structure of
planetary systems since the formation of the planets; some of these
processes might also
have played a crucial role in the accretion of the planets themselves.
The observational evidence that planetary systems can be very
different from each other, suggests that their dynamical evolutions
have been very diverse, probably as a result of a strong sensitivity 
of the dynamics on environmental parameters or initial conditions.

The goal of this chapter is to review our current understanding of the
possible dynamical evolutions of planetary systems and of the open
problems that we face. The main focus will be on our solar system,
because this is the system that we can best model, thanks to the vast
number of observational constraints. We will see that the 
dynamical history of our planetary system followed step-wise generic
processes, but was also characterized by a number of specific
``events'' that act like bifurcation points in the evolution of
planetary systems. I will then discuss what would have
happened if these events had occurred differently. This will give us some
guidance in understanding the origin of the diversity of planetary systems.

This chapter is divided in three parts. The first is devoted to the
early evolution of giant planets when they are still embedded in the
gas-disk, i.e. during the first few millions of years following the
formation of the central star. The second discusses the evolution of
the giant planets after the disappearance of the gas, when they
interact with a still massive planetesimal disk; this is the era of
debris disks, which are commonly observed around stars even as old as
1Gy. The third part will focus on terrestrial planets and on how their
accretion and evolution depend on the evolution of the giant planets
discussed before.

\section{The gas-disk era}
\label{gas}

\subsection{The formation of the giant planets}
\label{form}

There are two possible mechanisms by which we envision that giant
planets can form. The first is nicknamed the ``core-accretion
mechanism'': the coagulation of solid particles forms a core typically
of about 10 Earth masses ($M_\oplus$) while the gas is still present
in the proto-planetary disk; the core then traps by gravity a massive
atmosphere of hydrogen and helium from the disk (Pollack et al., 1996)
and becomes a giant planet. The second mechanism invokes the
gravitational instability of the gaseous component of the disk
(Cameron, 1978): a cold, massive proto-planetary disk can break into a
number of self-gravitating gas-clumps, which then contract forming
giant gaseous planets (Cassen et al., 1981; Boss, 2000, 2001, 2002;
see Durisen et al., 2007 for a review).

The debate to discriminate between these two models has been very
intense over the last 10 years. Now, several direct or indirect
observations suggest that the core-accretion mechanism is predominant
for the formation of the planets detected so far. First, interior
structure models of the giant planets of the solar system predict that
all of them have massive solid cores (Guillot, 2005; Militzer and
Hubbard, 2009; however see Nettelmann et al., 2008 for a model arguing
for a core-less Jupiter). Second, there is a clear correlation between
the metallicity of stars and the probability that said stars have
giant planets around them (Fisher and Valenti, 2005). Third,
transiting extra-solar planets are inferred to have solid cores whose
relative mass is correlated with the metallicity of the host star
(Guillot et al., 2006). All these features suggest that solids have a
crucial role in giant planet formation, an aspect that is difficult
to explain in the framework of the gravitational instability model
(Boss, 2002). Moreover, new hydro-dynamical simulations which model
more accurately the thermodynamics in the proto-planetary disk find
that formation of long-lived self-gravitating clumps of gas is likely
only at large distances from the central star ($>50$--100~AU; Boley,
2009). It is still unclear, though, whether the end-products of these
clumps can be giant planets or must be brown-dwarf-mass objects
(Stamatellos and Whitworth, 2008).

Thus, there is a growing consensus that the giant planets observed
within a few AUs from their parent stars formed by the core-accretion
process. The planets found at large distances from their parent stars
(for instance around HR 8799 - Marois et al., 2008- or Fomalhaut -
Kalas et al., 2008) are the best candidates to be the outcome of the
gravitational instability process. There is still a possibility,
though, that they are giant planets formed closer to the
star by core-accretion, which subsequently achieved large orbital
distances through planet-planet scattering (Veras et al., 2009) or
outwards migration (Crida et al., 2009). I will return to this,
further down in the chapter.

The core-accretion model, nevertheless, has its own problems. The main
difficulty is to understand how a $\sim 10 M_\oplus$ core could form
within a few million years (which is the typical survival time of a gas disk;
Haisch et al., 2001). In the classical view, these cores form by
collisional coagulation from a disk of planetesimals (small bodies of
sizes and compositions similar to current asteroids or comets).  In
this environment, gravity starts to play a fundamental role, bending the
trajectories of the colliding objects; this leads to an effective increase
of the collisional cross-section of the bodies by the so-called {\it
gravitational focussing factor} (Greenzweig and Lissauer, 1992). At
the beginning, if the planetesimal disk is dynamically very cold
(i.e. the orbits have tiny eccentricities and inclinations), the
dispersion velocity of the planetesimals $v_{\rm rel}$ may be smaller
than the escape velocity of the planetesimals themselves. In this
case, a process of {\it runaway growth} begins, in which the relative
mass growth of each object is an increasing function of its own mass
$M$, namely:
$$ 
{1\over
M} {{{\rm d}M}\over{{\rm d}t}} \sim {{M^{1/3}}\over{v_{\rm rel}^2}}\ ,
$$ 
(Greenberg et al., 1978; Wetherill and Stewart, 1989). However, as
growth proceeds, the disk becomes dynamically heated by the scattering
action of the largest bodies. When $v_{\rm rel}$ becomes of the order
of the escape velocity from the most massive objects (i.e. $v_{\rm
rel}\propto M_{\rm big}^{1/3}$), the runaway growth phase ends and
the accretion proceeds in an {\it oligarchic growth} mode, in which
the relative mass growth of the largest objects ${\rm d}\log M_{\rm
big}/{\rm d}t$ is proportional to $M_{\rm big}^{-1/3}$ (Ida and
Makino, 1993; Kokubo and Ida, 1998).

In principle, the combination of runaway and oligarchic growths should
continue until the largest objects achieve an {\it isolation mass},
which is a substantial fraction of the initial total mass of local
solids.  In the outer Solar System, beyond the so-called {\it
snowline}\footnote{The orbital radius beyond which temperature is cold
enough that water condenses into ice. The snowline is situated at
about 3-5 AU from a solar-mass star, depending on time and on disk
models (Min et al., 2011).} (Podolak and Zucker, 2004), if the initial disk is
sufficiently massive (about 10 times the so-called Minimal Mass Solar
Nebula or MMSN; Weidenschilling, 1977; Hayashi, 1981) it is expected
that the end result is the formation of a few super-Earths (Thommes et
al., 2003; Goldreich et al., 2004; Chambers, 2006), as required in the
core-accretion model for giant planet formation. $N$-body simulations,
though, show that reality is not so simple. When the cores achieve a
mass of about 1~$M_\oplus$ they start to scatter the planetesimals
away from their neighborhood, instead of accreting them (Ida and
Makino, 1993; Levison et al. 2010), which slows their accretion rate
significantly.  It has been proposed that gas drag (Wetherill and
Stewart, 1989) or mutual inelastic collisions (Goldreich et al., 2004)
prevent the dispersion of the planetesimals by damping their orbital
eccentricities, but in this case the cores open gaps in the
planetesimal disk (Levison and Morbidelli, 2007; Levison et al.,
2010), like the satellites Pan and Daphis open gaps in Saturn's
rings.  Thus the cores isolate themselves from the disk of solids.
This effectively stops their growth.  It has been argued that planet
migration (Alibert et al., 2004) or the radial drift of small
planetesimals due to gas drag (Rafikov, 2004) break the isolation of
the cores from the disk of solids but, again, $N$-body simulations
show that the relative drift of planetesimals and cores simply
collects the former in resonances with the latter (Levison et al.,
2010); this prevents the planetesimals from being accreted by the cores.

In summary, the accretion of massive cores is still an open problem,
and I am suspicious of synthetic models where simple formul\ae\ for
the mass-growth of the cores are made up to achieve the desired
result, without any correspondence with the outcomes of $N$-body
simulations (which should be considered as a sort of ``ground truth''
for dynamical processes).

Another problem of the core-accretion model originates from Type-I
migration.  Type-I migration is the label denoting the radial drift of
planetary cores (with masses ranging from that of Mars to several Earth
masses) due to their gravitational interaction with the gaseous
component of the disk. Analytic and numerical studies have shown that
a planetary core generates a spiral density wave in the disk
(Goldreich and Tremaine,1979,1980; Ward, 1986, 1997; Tanaka et al. 2002). In
the outer part of the disk, the wave trails the core.  Thus, the
gravitational attraction that the wave exerts on the core results in a
negative torque that slows the core down. In the inner part of the
disk, the wave leads the core, and therefore it exerts on it an
acceleration torque. The net effect on the core depends on the balance
between these two torques of opposite signs. Ward (1997) showed that
in general cases, i.e. for disks with power-law radial density
profiles, the negative torque exerted by the wave in the outer disk
wins. Consequently the core has to lose angular momentum, and its
orbit shrinks: the planetary core migrates towards the central star,
with a speed:
$$
da/dt \propto M_p \Sigma_g (a/H)^2\ ,		
$$ 
where $a$ is the orbital radius of the planet, $M_p$ is its mass,
$\Sigma_g$ is the surface density of the gas disk and $H$ is its height
at the distance $a$ from the central star.

Precise calculations show that an Earth-mass body at 1 AU, in a
MMSN  with scale height $H/a$ = 5\%, migrates into
the star in 200,000y. Thus, planetary cores should fall onto the
central star well before they can attain the mass required to
capture a massive atmosphere and become giant planets. 

Several mechanisms that might weaken or prevent Type-I migration have
been investigated. First, turbulence may turn inward Type-I migration
into a random walk (Nelson and Papaloizou, 2003; Nelson, 2005), which
could save at least some of the cores. Second, if there is a steep
positive $d\Sigma_g/da$ at some location in the disk, for instance at
the outer edge of a partially depleted central cavity, inward Type-I
migration should stop there (Masset et el. 2006). Finally, it has been
recently shown that migration can be outwards in the inner part of the
disk, which transports and dissipates heat inefficiently due to its
large opacity (Paardekooper and Mallema, 2006; Baruteau and Masset,
2008; Kley and Crida, 2008; Paardekooper et al., 2010). In this case,
all cores would migrate towards an intermediate region of the disk,
where Type-I migration is effectively erased (Lyra et al., 2010).

A particularly puzzling aspect of the core-accretion process is the
evidence that in our solar system massive cores of $\sim 10 M_\oplus$
formed in the giant planets region, whereas in the inner solar system
the planetary embryos resulting from the runaway/oligarchic growth
process presumably had masses smaller than the mass of Mars (Wetherill
and Stewart, 1993; Wetherill, 1992; Weidenschilling et al.,
1997). This jump of two orders of magnitude in the masses of
embryos/cores from the inner to the outer solar system is difficult to
understand. In fact, the surface density of solids in the disk should
have had a 'jump' at the snowline of only a factor of $\sim 2$, given
the revised solar C/O abundance (Lodders, 2003). Moreover, the orbital
frequency (that sets the speed of all dynamical processes, including
accretion) decreases with increasing distance from the star. So, what
makes the outer solar system so favorable for the formation of massive
cores?  To answer this question, several investigators searched for
mechanisms that can concentrate a large amount of solids (well above
the initial surface density) in some localized region of the disk, so
to achieve a sweet spot for the formation of a massive object. Some
proposed mechanisms still give a pivotal role to the snowline (Morfill
and Volk, 1984; Ida and Lin, 2008), but others are based on the
concentration of boulders in long-lived vortices (Barge and Sommeria,
1995; Lyra et al., 2009a, 2009b), or on halting migration of planetary
embryos at a given orbital radius (Masset et al., 2006; Morbidelli et
al., 2008a; Paardekooper and Papaloizou, 2009; Sandor et al, 2011),
which are independent of the snowline location.  Obviously more work
is needed to understand which mechanism is relevant and dominant.
Depending on future results, it might turn out that the wide-spread
expectation that giant planets form by the core-accretion mechanism
only beyond the snowline is naive; instead, some giant planets might
have formed in the warmer regions of the disk (Bodenheimer et al.,
2000).

\subsection{Once giant planets are formed: Type II migration and its
  consequences}
\label{TypeII}

By the action-reaction principle, the torques exerted by the disk onto
the planet are symmetrically exerted by the planet onto the
disk. Thus, the planet exerts a positive torque onto the outer disk,
i.e. it pushes the outer disk outwards, while it exerts a negative
torque onto the inner disk, i.e. it pushes it inwards.  For small mass
objects, such as the planetary cores considered in the previous
section, these torques are overcome by the internal torques of the
disk, due to viscosity and pressure, so that the mass distribution in
the disk -averaged over the azimuthal coordinate- is not significantly
affected by the presence of the planet. However, this is no longer
true if the planet is sufficiently massive (several tens of Earth
masses for typical values of the disk's parameters); in this case the
internal torques of the disk cannot oppose the torques suffered from
the planet. Consequently, the inner and outer parts of the disk are
effectively repelled, and a gap in the gas distribution opens around
the planet's orbit (Lin and Papaloizou, 1986a; Crida et al., 2006).

Once a giant planet has opened a gap in the disk, it is condemned to
stay in the middle of the gap. In fact, if it approached the inner
edge of the gap, the distance of the planet from the inner disk would
decrease and that from the outer disk would increase, so that the
torque felt from the inner disk would become stronger than that felt
from the outer disk. Thus the planet would be pushed back towards the
center of the gap. The symmetric situation would occur if the planet
approached the outer edge of the gap. Consequently, any radial
migration of the planet has to follow the radial migration of the gap.
The global evolution of the disk occurs on a viscous timescale
$T_\nu=a^2/\nu$, where $\nu$ is the disk viscosity (Lynden-Bell and
Pringle, 1974). So, the radial displacement of the gap and the
migration of the planet have to occur on this timescale (Lin and
Papaloizou, 1986b). Unless the planet is close to the outer edge of the
disk (Veras and Armitage, 2004), the planet has to migrate towards the
star, because this is the natural direction of evolution of an
accretion disk (Lynden-Bell and Pringle, 1974; Lin and Papaloizou,
1986b). The migration of a planet associated with a gap is called ``Type II
migration''.

Type II migration explains in a natural way the presence of giant
planets on orbits very close to the parent stars (Lin et al., 1996),
which is a very common characteristic in the population of extra-solar planets
discovered to date. However, in our solar system the giant planets have
orbital radii of several AUs, comparable to the orbital radii
at which said planets are expected to have formed (see
sect.~\ref{form}). Moreover, several extra-solar planets have also been
discovered on orbits with semi major axes larger than $\sim 3$ AU. What
happened in these cases? Why is Type II migration sometimes
ineffective?

\begin{figure}[t!]
\centerline{\includegraphics[height=6.cm]{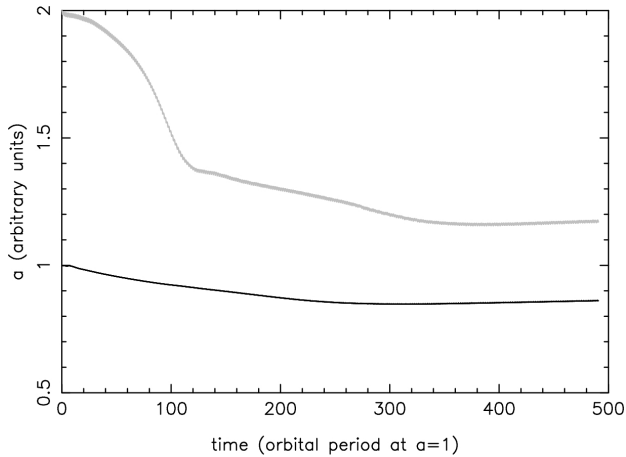}}
\vspace*{-.3cm} 
\caption{\small An illustration of the dynamical evolution of Jupiter
  and Saturn in the gas-disk, as in Masset and Snellgrove (2001). The
  black and grey curves show the evolutions of the semi major axes of
  Jupiter and Saturn, respectively. Capture in the 2/3 mean motion
  resonance occurs when the migration of Saturn is reversed.}
\label{MS01} 
\end{figure}

For our solar system, the key to answer this question seems to be the
co-existence of Jupiter and Saturn, with their specific mass ratio. In
fact, hydro-dynamical simulations, where Jupiter and Saturn are
simultaneously taken into account with fixed masses, show that Saturn
migrates the fastest, due to it smaller mass; therefore, it has no
trouble in approaching Jupiter until the two planets are caught in
resonance (Masset and Snellgrove, 2001). In disks with mass comparable
to the MMSN, the most likely end-state is the capture in the mutual
2/3 mean motion resonance, where the orbital period of Saturn is 1.5
that of Jupiter. This occurs even if Saturn is initially beyond the
1/2 resonance or locked into the 1/2 resonance (Pierens and Nelson,
2008). Stable capture into the 1/2 resonance is possible only for
disks with surface density decaying less steeply than 1/r or in
low-mass disks (Zhang and Zhou, 2010). Once locked in the 2/3
resonance, the inward migration of Jupiter and Saturn stops (Masset
and Snellgrove, 2001; Morbidelli and Crida, 2007), which explains why
Jupiter did not migrate all the way close to the Sun. These results
hold both for disks with constant viscosity and for the so-called
$\alpha$-disks (Shakura and Surayev, 1973) and do not depend
critically on the value of the viscosity.

However, things may not be so simple in reality. All the
hydrodynamical simulations that show that Saturn captures Jupiter in
resonance, assume fixed masses for the planets. A natural question
arises: is it still reasonable to expect resonance capture if the
migration histories of the planets are coupled with their accretion
histories?  At first sight the answer is negative. If the second
planet forms later than the first one, its migration history should
just replicate that of the first planet, but later in time. More
simply, the second planet should always lag behind the first one, as
it is just repeating the evolution of the first planet, just at a
later time. Thus, it appears that Saturn could catch Jupiter in
resonance only if the accretion histories of the two planets were
different. In particular, if Jupiter grew very rapidly to its current
mass, it would have passed very quickly to a Type-II migration mode,
which is relatively slow.  Instead, if Saturn grew more gradually than
Jupiter and spent more time near a Saturn-mass, it would have
undergone fast migration for a longer period and hence could have trapped
Jupiter in resonance. It is unclear why Saturn should have grown more
slowly than Jupiter. Possibly, the opacity of the disk increased from
the time of accretion of Jupiter to that of Saturn, thus slowing down
the gas-accretion rate onto the planet.

Interestingly, among the collection of extra-solar planets, we see at
least three systems where the Saturn-analog did not capture in resonance the
Jupiter-analog (HD 12661, HD 13498, HIP 14810). But in
these cases, the inner, more massive planet is closer than 1 AU to the
star and the outer, lighter planet is more than three times further
away. This is obviously very different from the orbital architecture
of Jupiter and Saturn, or of the system OGLE-06-109L (which is a sort
of twin of the Jupiter-Saturn system), which suggests that a different
evolution occurred in these cases.  I argue that the capture in
resonance between a Jupiter-analog and a Saturn-analog is an event that
may or may not happen, according to the accretion histories of these
planets; depending on this binary possibility, the systems evolve along
clearly different paths.  

Let's admit now that Jupiter and Saturn got captured in their mutual
2/3 resonance.  Morbidelli and Crida (2007) showed that the subsequent
dynamical evolution of these planets depends on the properties of the
disk, particularly the scale height. For thick disks (about 6\% in
scale height for a typical viscosity), the migration is very slow, and
the planets remain at effectively constant distance from the central
star. But for disks with decreasing thickness, outward migration
becomes increasingly fast. In principle, in thin disks outward
migration can bring the planets up to ten times further than their initial
location in a few thousands orbital periods (Crida et al., 2009),
which might explain the orbits of some of the planets discovered by
direct imaging beyond several tens of AUs from their parent
stars. Until now, outward migration of Jupiter and Saturn from inside
$\sim 4$ AU was considered incompatible with the existence of the
asteroid belt - therefore only proto-Solar disk models which
prohibited outward migration were considered viable (Morbidelli et al.,
2007). However, as shown by Walsh et al. (2011) and discussed in
Sect.~\ref{GT}, this is not true, releasing the constraints against
Jupiter's outward migration. This opens a new degree of freedom to
model the evolution of the solar system, as it will be shown at the
end of this chapter. 

For two giant planets to avoid inward migration as discussed above, it
is essential that the mass of the outer planet is a fraction of the
mass of the inner planet, as in the Jupiter-Saturn case (Masset and
Snellgrove, 2001; Morbidelli and Crida, 2007). I provide here an
heuristic, but intuitive explanation of this statement. When two giant
planets are close enough to each other, they evolve inside a common
gap of the gas-density distribution. The inner planet, being closer to
the inner edge than to the outer edge of the common gap, feels a
positive torque and would tend to migrate outwards; the outer planet,
being closer to the outer edge of the common gap, feels a negative
torque and would tend to migrate inwards.  If the planets are locked
in resonance their relative orbital separation cannot change (if they
are not yet in resonance they move towards each other until they are
captured and locked into a resonance). Thus, the direction of 
migration of the pair of planets depends on which of the two torques
dominates. Each torque is proportional to the surface density of the
disk adjacent to the planet and to the square of the mass of the
planet itself (Goldreich and Tremaine, 1979). Because the planets
deplete partially the disk in the region between the star and their
innermost orbit (a partial cavity; Crida and Morbidelli, 2007), the    
surface density in the outer disk is typically larger than in the
inner disk; so, a necessary condition to avoid inward migration
(i.e. to make the torque felt by the inner planet larger than that
felt by the outer planet) is that the inner planet is more massive.

Therefore, this mechanism predicts that no pair of resonant giant
planets with narrow orbital separation, with the outer planet
significantly lighter than the inner one, should ever be found close
to the parent star. In fact, planets in this configuration should have
avoided inward migration. So far, this prediction is validated by
observation. I fact, there are many pairs of planets in resonance (or
close to), near their star, but none of these cases exhibits a
Jupiter/Saturn mass ratio. The absence of this configuration, which is
statistically significant if one assumes that the mass ratio should be
random, strongly supports the theoretical result that resonant planets
in Jupiter/Saturn mass ratio move outwards, and therefore cannot be
found within the range of stellar distances that can be probed by
radial velocity observations (OGLE-06-109L system was in fact
discovered by micro-lensing).

There is an intriguing aspect in this view of the evolution of Jupiter
and Saturn. When the two planets do not migrate inwards at the nominal
Type II migration rate, there is necessarily an inward flow of gas,
from the outer disk to the inner disk, through the common gap.  Thus,
the outer planet should presumably accrete more of the incoming
material, narrowing the mass difference with the inner planet. This
raises the question of why Saturn remained smaller than Jupiter.  The
answer may be that the gas disk was rapidly disappearing while Jupiter
and Saturn were undergoing the dynamical evolution described above, so
that Saturn failed to grow further. There are several lines of
evidence in favor of a formation of Jupiter and Saturn in a
dissipating disk. First, Jupiter's atmosphere is enriched in elements
heavier than helium by a factor of 3-4 relative to solar composition
(Wong et al., 2004), while Saturn is enriched by a factor 11 in Carbon
(Fouchet et al., 2009). Guillot and Hueso (2006) have argued that the
easiest explanation for this fact is that hydrogen and helium had been
already depleted by a factor of 3-4 in the disk by the time Jupiter
captured its atmosphere (and, following this logic, by a factor of 11
by the time Saturn captured its atmosphere).  Second, the favored
model for the accretion of the regular satellites of the giant planets
also requires that said satellites formed in gas-poor circum-planetary
disks (Canup and Ward, 2006). Third, the common explanation for why
Uranus and Neptune failed to accrete massive atmospheres is that the
gas disappeared before they had a chance to do so (Pollack et al.,
1996). All these arguments suggest that the 4 giant planets formed in
a temporal sequence, from Jupiter to Saturn and then to Uranus and
Neptune, while the disk was being dispersed. Finally, recall that the
Solar composition is at the low-end of the metallicity range for
planet-bearing stars. In the core-accretion model of giant planets,
the metallicity of the star is correlated with the speed of accretion
of the cores, so that stars that are too poor in metals fail to form
giant planets before the disappearance of the disk (Ida and Lin,
2004). This suggests that the solar system barely made its giant
planets, while the disk was being dispersed.

\begin{figure}[t!]
\centerline{\includegraphics[height=6.cm]{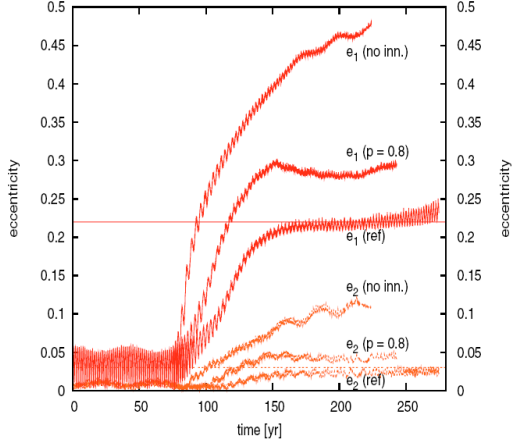}}
\vspace*{-.3cm} 
\caption{\small The evolution of the eccentricities of planets {\it b}
  and {\it c} around GJ876 during their putative inward resonant
  migration. The eccentricity of the interior planet is labelled
  $e_1$, that of the exterior planet $e_2$. The evolutions labelled
  ``no inn.'' assume that no disk is present inside the orbit of the
  interior planet; in this case the eccentricities seem to grow
  indefinitely. The evolutions labelled ``ref.'' and ``p=0.8'' account
  for an inner disk, and differ just for the value of a technical
  simulation parameter. In these cases the eccentricities attain
  equilibrium values. In the ``ref'' case, the final eccentricities
  reproduce the eccentricities inferred from observations (horizontal
  lines). From Crida et al., 2008.}
\label{Crida_e} 
\end{figure}

If this explanation may be satisfying for our solar system, it is
nevertheless interesting to discuss what would have happened if
Jupiter and Saturn had formed earlier, when the disk was still
massive. The issue is not only academic, as it can be pertinent for systems
with a higher initial metal content which, as suggested above,
should form planets faster.  In a gas rich disk, Saturn would have
eventually become as massive, or even more massive, than
Jupiter. Thus, the two planets would have resumed an inward migration
(Morbidelli and Crida, 2007).  The orbital eccentricity of two planets
migrating in resonance tends to increase monotonically (Ferraz-Mello
et al., 2003; Kley et al., 2004). This effect is contrasted by the
action of the disk, which tends to damp the planets' eccentricities
(Kley and Dirksen, 2006). A crucial role is played by the disk inside
the orbit of the inner planet. This inner disk tends to be partially
depleted due to the presence of the planet(s). The level of depletion
depends on several parameters such as the viscosity of the disk, its
scale height, its inner radius, the mass of the planet(s) relative to
the disk etc. (Crida and Morbidelli, 2007).  If the inner disk is
depleted substantially, the damping effect on the eccentricity of the
inner planet is strongly reduced. In this situation, the eccentricity
of the inner planet keeps growing, until the pair of planets becomes
dynamically unstable and mutual close encounters are triggered (Kley
et al., 2005). This may be the case of many, if not most, of the
planetary systems. In fact, mutual scattering seems to be the major
mechanism responsible for the eccentricity distribution observed in
the extra-solar planets collection (see sect.~\ref{scatter} for a
more complete discussion).  However, if the inner disk is not very
depleted, the eccentricities of the planets grow until a limit value
is achieved (Crida et al., 2008; see Fig.~\ref{Crida_e}). This process
can leave the planets at the disappearance of the disk on stable
resonant orbits with moderate eccentricities and thus it can explain
the pairs of planets in resonance observed to date.

In conclusion, we have seen in this section the first crucial
``bifurcations'' in planetary evolution that can account for at
least part of the great diversity observed in planetary systems. In
fact, assuming that giant planets form in sequence at increasing
distances from the central star, most of the observed diversity of
planetary systems could stem from the occurrence or avoidance of two
events: (i) the capture in resonance of the first, inner planet by the
second, initially smaller one, which stops inward migration (often
triggering outward migration) and (ii) the growth of the outer planet
beyond the mass of the inner one, which causes inward migration of
both planets to resume. The Solar System structure results from the
occurrence of (i) and avoidance of (ii).  Systems like HD 12661, with
a close-in massive planet and a distant smaller planet, result from
the avoidance of (i). Resonant giant planets close to their stars,
like those in the GJ876 system, result from the occurrence of both (i)
and (ii). Unstable systems, ultimately leaving behind one giant planet
on an eccentric orbit, may also result from the occurrence of both (i)
and (ii), but in cases where there was not enough eccentricity damping
because of a depleted inner disk.

Here, the case with two planets was the only one considered;
obviously, the tree of possible evolutions can only become more
complicated if more giant planets are involved, and the final outcomes
can be even more diverse.

\subsection{Planet-planet scattering as the
dominant orbital excitation process} 
\label{scatter}

One of the greatest surprises that came with the discovery of
extra-solar planets is the realization that most planets have orbital 
eccentricities much larger than those characterizing the planets of
our solar system. Eccentricities of about 0.4 are quite common in the
extra-solar planets collection; some planets have eccentricities
larger than 0.6, values that in our solar system are common only for
comets! 

Whatever the preferred model of giant planet formation (core-accretion
or gravitational instability; see sect.~\ref{form}), it is expected
that planets have originally small orbital eccentricities, because
they form from a circum-stellar disk whose streamlines are basically
circular.  There has been a lively debate on whether subsequent
planet-disk interactions can raise the planets' eccentricities up to
the observed values. Papaloizou et al., (2001) concluded, with
numerical experiments and theoretical considerations, that
eccentricity growth is not possible for planetary masses below 10-20
Jupiter masses. Instead, Goldreich and Sari (2003) argued with
theoretical considerations that, under some conditions, depending
mostly on disk's thickness, giant planets of more moderate masses
could have their orbital eccentricity excited, although they could not
estimate the magnitude of this excitation. More recent hydro-dynamical
simulations (D'Angelo et al., 2006; Kley and Dirksen, 2006) showed
that planets with masses larger than $\sim $2--3 Jupiter masses, under
some conditions, can have eccentricities excited by the disk, but only
to moderate values ($\sim 0.1$--0.2), definitely lower than those
characterizing many, if not most, of the extra-solar planets. In most
cases, the planet-disks interactions rather seem to lead to
eccentricity damping. Thus, a more generic orbital excitation
mechanism seems to be required to explain the observations.

\begin{figure}[t!]
\centerline{\includegraphics[height=6.cm]{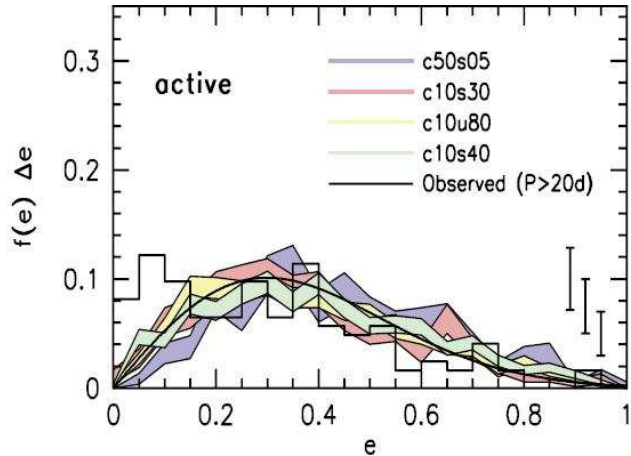}}
\vspace*{-.3cm} 
\caption{\small Final eccentricity distribution of simulated ensembles
  of planetary systems that underwent a dynamical instability sometime during
  the full simulation time-span of $10^8$~y. The color bands correspond 
  to different ensembles, characterized by different initial
  conditions.  The histogram shows the observed eccentricity
  distribution of extra-solar planets with orbital period longer than
  20~d, according to Butler et al. (2006). The observed distribution
  and the final distributions resulting from the simulations agree
  very well, with the exception of an excess of observed planets with
  $e<0.2$. This is probably due to planets that never underwent a
  significant dynamical instability. From Juric and Tremaine (2008).}
\label{Juric} 
\end{figure}

Soon after the discovery of the first eccentric planets, it was
pointed out that mutual encounters between planets can easily provide
strong orbital excitation (Rasio and Ford, 1996; Weidenschilling and
Marzari, 1996; Lin and Ida, 1997; Levison et al., 1998; Ford et al.,
2001; Marzari and Weidenschilling, 2002; Adams and Laughlin,
2003). More recent studies (Juric and Tremaine, 2008; Chatterjee et
al., 2008; Ford and Rasio, 2008; Raymond et al., 2009) show that
random systems of giant planets initially on unstable, quasi-circular
orbits evolve through close encounters until a dynamical relaxation
state is achieved with the ejection or collision of some planets. In
these models the final eccentricity distribution of the surviving
planets is remarkably similar to that of known extra-solar planets
(see Fig.~\ref{Juric}). Moreover, Juric and Tremaine (2008) and
Raymond et al. (2009) showed that the final orbital spacing of the
surviving planets is also in good agreement with the observations of
extra-solar systems of two or more non-resonant planets. These systems
typically look ``packed'', in the sense that the orbital separation
(apocenter to pericenter) between neighboring planets is not much
larger than what is required by the Hill-stability criterion (Barnes
and Greenberg, 2006).  All these quantitative results give strong
support to the idea that planet-planet scattering is the main
mechanism sculpting the orbital distribution of extra-solar
planets. In addition, Veras et al. (2009) pointed out that
planet-planet scattering can propel a planet in the region beyond
100~AU, which can explain some of the extra-solar planets imaged at
large distances from their parent stars\footnote{In summary, three
mechanisms have been proposed to move planets from the snowline region
to large distances from the central star: (i) outward Type-II
migration of planets originally formed in the outer part of a disk in
rapid viscous spreading (Veras and Armitage, 2004); (ii) outward
migration of a pair of resonant planets with a Jupiter-Saturn mass
hierarchy (Crida et al., 2009); (iii) scattering of a planet to a wide
elliptic orbit (Veras et al., 2009). These mechanisms are potential
alternatives to the possibility that distant planets formed in-situ,
by the gravitational instability mechanism (Boley et al., 2009)}.

\begin{figure}[t!]
\centerline{\includegraphics[height=6.cm]{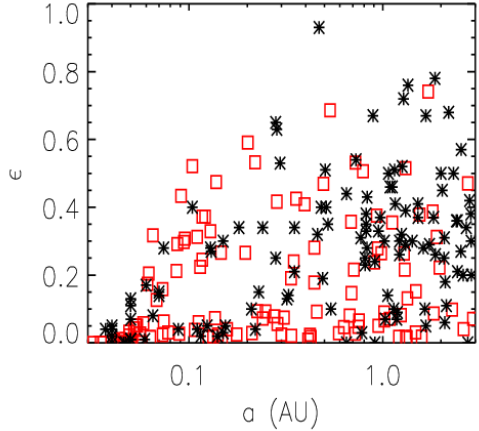}}
\vspace*{-.3cm} 
\caption{\small The final semi major axis
vs. eccentricity distribution of extra-solar planets (squares) in the
simulations of Moorhead and Adams (2005) which account for (i) inward
migration, (ii) eccentricity damping due to the disk (with an assumed
timescale of 0.3~My) (iii) tidal circularization and (iv) radial
velocity detection biases. The stars show the distribution of the
extra-solar planets known at the time.}
\label{Althea} 
\end{figure}

It is important to remember that, assuming giant planets formed beyond
an hypothetical snowline at 3--5 AU, the planet-planet scattering
mechanism alone cannot explain the observed semi major axis
distribution of extra-solar planets (see for instance Marzari and
Weidenschilling, 2002). In fact, through scattering events, planets
can have difficulty reaching orbits with semi major axis smaller than half of
the initial value of the inner planet, with the exception of those
objects scattered onto orbits with very large eccentricities and small
periastron distances, which may then become circularized by tidal
dissipation. Thus, migration is still required to explain the presence
of a large number of planets on orbits with small semi major
axes. Adams and Laughlin (2003) and Moorhead and Adams (2005) modeled
the interplay between migration and scattering. They used N-body
simulations, with fictitious forces to mimic the effect of the disk on
the planets, for what concerns both semi major axis decay and
eccentricity damping. After tuning a few parameters (the lifetime of
the disk, the timescale of the eccentricity damping, etc.)  and
accounting for observational biases, Moorhead and Adams obtained a
very good reproduction of the two-dimensional $(a,e)$ distribution of
the exoplanets detected by the radial velocity technique
(Fig.~\ref{Althea}). Although very appealing, this result may be
questioned because the numerical recipes used to mimic migration and
damping have been based on simple analytic estimates.  Reality may be
more complicated. For instance, as we have seen above, a system of two
planets in resonance may migrate at a very different speed than a
single planet in Type II migration (Morbidelli and Crida, 2007); the 
disk may not always damp the eccentricity of a planet, but could also
sustain it, depending on the planet mass and initial eccentricity
(Kley and Dirksen, 2006); migration direction could be reversed for
eccentric planets (D'angelo et al., 2006); finally mass accretion onto
the planets is neglected in the Moorhead and Adams model.

For all these reasons, I think that two key questions have not yet
been answered in a definitive way: why do giant planets become unstable in
the first place? And when do they become unstable, relative to the gas
disk lifetime? I speculate a bit on both issues below. 

Concerning the first question, there are in principle two answers:
either planetary systems become unstable because the planets grow in
mass, or because the planets are brought too close to each other by migration
processes. I would tend to exclude the first case as a dominant
mechanism to explain the large eccentricities of extra-solar giant
planets, for the following reason. Imagine a system of cores close to
each other (maybe brought by Type I migration into mutual resonances
of type $n/(n+1)$, with quite large $n$). Given that the time required
to trigger the runaway accretion of a massive atmosphere is much
longer than the accretion of the atmosphere itself (Pollak et al.,
1996), it is unlikely that all the cores would accrete massive
atmospheres and become giant planets simultaneously. More
realistically one core would start the accretion of the atmosphere
first, and it would acquire a mass much larger than those of the other
planets at that time. The system would become unstable, because it is
too closely packed to stand the newly born giant planet.  However, the
scattering phase would bring the cores onto orbits with large
eccentricities but it would leave the newly born giant planet onto a
quasi-circular orbit: this does not correspond to the orbital
distribution of the extra-solar planets that we see. Moreover, the
orbital eccentricities of the cores would be damped very fast by the
disk (Cresswell et al., 2007; Bitsch and Kley, 2010) and, by the time
these cores become in turn giant planets, they would be again on
quasi-circular orbits. I'm not denying that this kind of evolution can
occur. A system of multiple giant planets probably forms through
several repetitions of the events described above\footnote{The reader
should remember that two planets are stable if their orbital
separation is a few times their mutual Hill radius
$R_H=\bar{a}[(m_1+m_2)/M_S]^{1/3}$, where $m_1$ and $m_2$ are the
masses of the two planets and $\bar{a}$ is their mean semi major axis,
while $M_S$ is the mass of the star. Suppose now that planets tend to
acquire orbits whose mutual separation is not much larger than this
Hill-stability limit. If a planetary system is made of two planetary
cores, when one of two objects becomes a giant planet, the system is
likely to be destabilized because $R_H$ increases by a factor 3--4 as
the mass of one planet grows by a factor 30--60. Instead, a stable
system made of one giant planet and one core is not likely to be
strongly destabilized when the core grows to the status of a giant
planet, because $R_H$ increases only by a factor $\sim
2^{(1/3)}=1.25$. Similarly, in a system made of one giant planet and
two cores, the growth of one of the cores to the status of a giant
planet is likely to destabilize the remaining core, but not the first
planet.}. My claim is that this is probably not the process that
sculpts the {\it final} orbital distribution of giant planet
systems. I think that a more generic mechanism to achieve the ultimate
dynamical instability (the one responsible for the final orbits) is
that two or more giant planets, once fully formed, are brought in
resonance with each other by Type II migration, which causes a
subsequent increase of their orbital eccentricities (Ferraz-Mello et
al., 2003; Kley et al., 2004, 2005; Crida et al., 2008).

\begin{figure}[t!]
\centerline{\psfig{figure=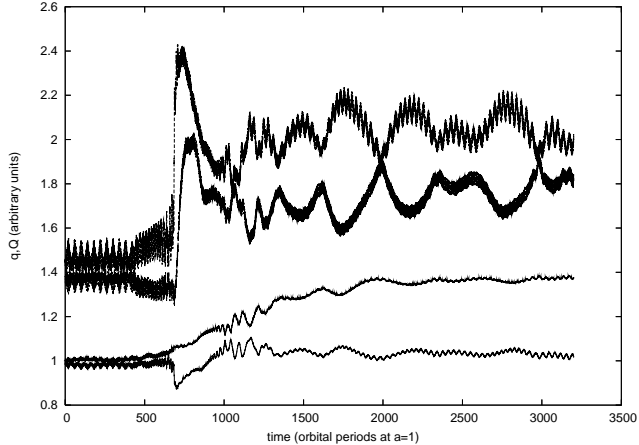,height=6cm,angle=-90}}
\vspace*{-.3cm} 
\caption{\small The evolution of a pair of giant planets initially on
  unstable orbits. For each planet, the pair of curves shows the
  evolution of the periastron ($q$) and apoastron ($Q$)
  distances. Thus, when the two curves are close, the orbit is almost
  circular and $q\sim Q\sim a$, where $a$ is the orbital semi major
  axis. The light curves at the bottom are for the inner planet, with a 3
  Jupiter mass; the thick curves at the top are for the outer planet, with one
  Jupiter mass. Notice that the orbits become very eccentric and
  separate from each other at the time of the instability, after about
  500-700 orbits of the inner planet. Subsequently, the eccentricities
  of the planets are damped and the inward migration of the outer
  planet brings it in the 1/2 resonance with the inner one
  ($t=1,000$). Once in resonance, the eccentricity increases again and
  starts to have long-period oscillations. A relatively stable
  configuration is achieved. From Morbidelli and Crida (2007).
 }
\label{3J-J} 
\end{figure}

This brings us to the second question, on the timing of the
instability. In the Adams and Laughlin (2003) and Moorhead and Adams
(2005) models, the planets become unstable while they are still
migrating in the gas disk. However, our limited experience with
hydro-dynamical simulations of giant planets scattering each other in
gas-disks shows that the evolution can be quite different from the one
modeled in those works.  Scattered planets tend to acquire orbits
which are more separated in semi major axis and more eccentric, but
then the eccentricities are damped by the disk and the planets migrate
back into a new, more stable resonant configuration, with moderate
eccentricities (see for instance Fig.~\ref{3J-J} or Moekel et al.,
2008). It is possible that systems with more planets develop more
violent instabilities that extend also longer in time. However, new
simulations with three planets, presented in Marzari et al. (2010),
again show that the planets surviving at the end of the instability
phase have low-eccentricity orbits. Until we know more from
hydro-dynamical simulations about the dynamics of eccentric planets,
it is premature to conclude whether instabilities produced during the
gas-disk phase would lead to the observed orbital distribution of
extra-solar planets.  The other possibility is that during the gas
disk phase planets acquire stable resonant and eccentric orbital
configurations, which become unstable when the gas is removed. This is
the approach of Lin and Ida (1997), Levison et al. (1998), Juric and
Tremaine (2008), Chatterjee et al. (2008), to quote just a few works.

The facts that most extra-solar planets have quite large
eccentricities and that there are only a few pairs of stable resonant
planets, suggest that orbital instability in planetary systems is more
the rule than the exception. I think that it would be quite surprising
if most planet configurations achieved through migration were stable
in presence of gas and unstable in absence of gas. In fact, the gas
has some stabilizing effect due to the eccentricity damping that it
exerts, but it also drives migration, which in turn excites the
eccentricity.  The two effects cancel out when a limit eccentricity is
attained (Fig.~\ref{Crida_e}). At this point, the gas should not play
any longer any crucial role in maintaining stability. One possibility
is that most resonant configurations achieved through migration are
unstable in both cases (with gas and without gas), but the instability
manifests itself on timescales of several millions 
of years, i.e. well after that the gas has been removed. To have a
better appreciation of reality, we need more systematic
hydro-dynamical simulations of the dynamics of sets of giant planets
embedded in gas-disks, followed by the investigation of their
subsequent long-term evolutions after the gas removal. A work of this
kind has been done for the planets of our solar system (Morbidelli et
al., 2007; see sect~\ref{4plSS}), but it is obviously more demanding
in general, given the volume of parameter space that needs to be
explored.

Naively, I would have expected that planets that develop instabilities
and mutual encounters after having migrated to the vicinity of 
the central star acquire smaller eccentricities than those that do so
further away from the star. This is because the eccentricity acquired
in an encounter is proportional to the ratio between the velocity kick
received during said encounter and the orbital velocity. The former
depends on the escape velocities from the surface of the planets and
is independent of the orbital radius, while the latter is larger for
the close-in planets. Instead, Juric and Tremaine (2008) showed
numerically that the eccentricity distribution achieved at dynamical
relaxation (i.e. after many encounters) is essentially independent of
the semi major axes of the planets. Given that we do not see any clear
correlation between distance and eccentricity in the extra-solar
planets collection (apart from the tidal circularization zone) this
result gives a quite strong support to the idea that planets first
migrate towards the central star and then, after gas removal, become
unstable.

\subsection{A plausible evolution of the four giant planets of the
  solar system}
\label{4plSS}

I now come back to our solar system, to outline a plausible scenario
for the evolution of the 4 major planets during the gas-disk phase.
As we have seen above (sect.~\ref{TypeII}), hydro-dynamical
simulations strongly suggest that Jupiter and Saturn rapidly reached a
2/3 resonant orbital configuration and that this prevented them from
migrating further towards the Sun. Once in resonance, these giant
planets either remained on non-migrating orbits or migrated outwards.

\begin{figure}[t!]
\centerline{\includegraphics[height=8.cm]{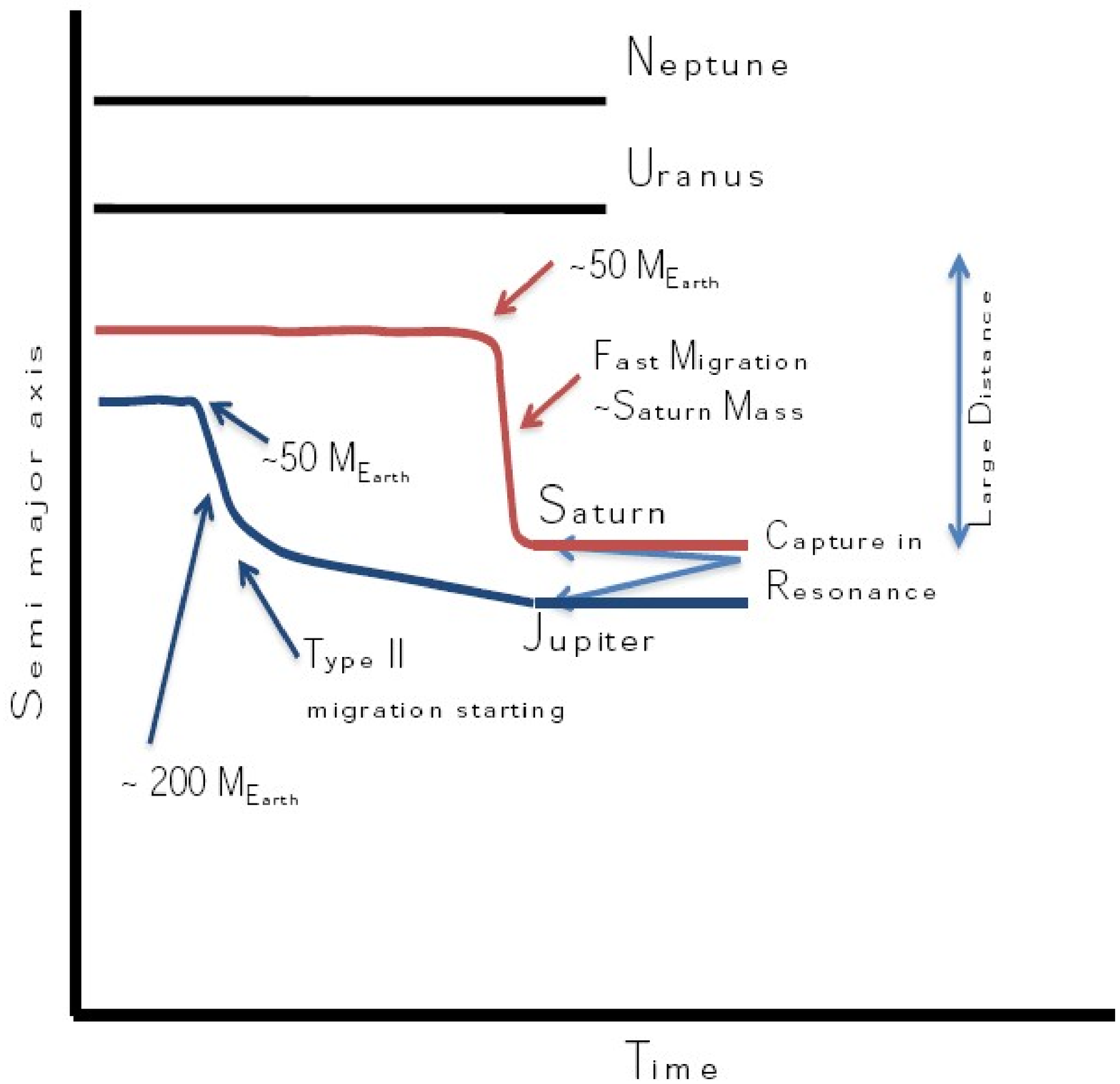}\quad\includegraphics[height=8.cm]{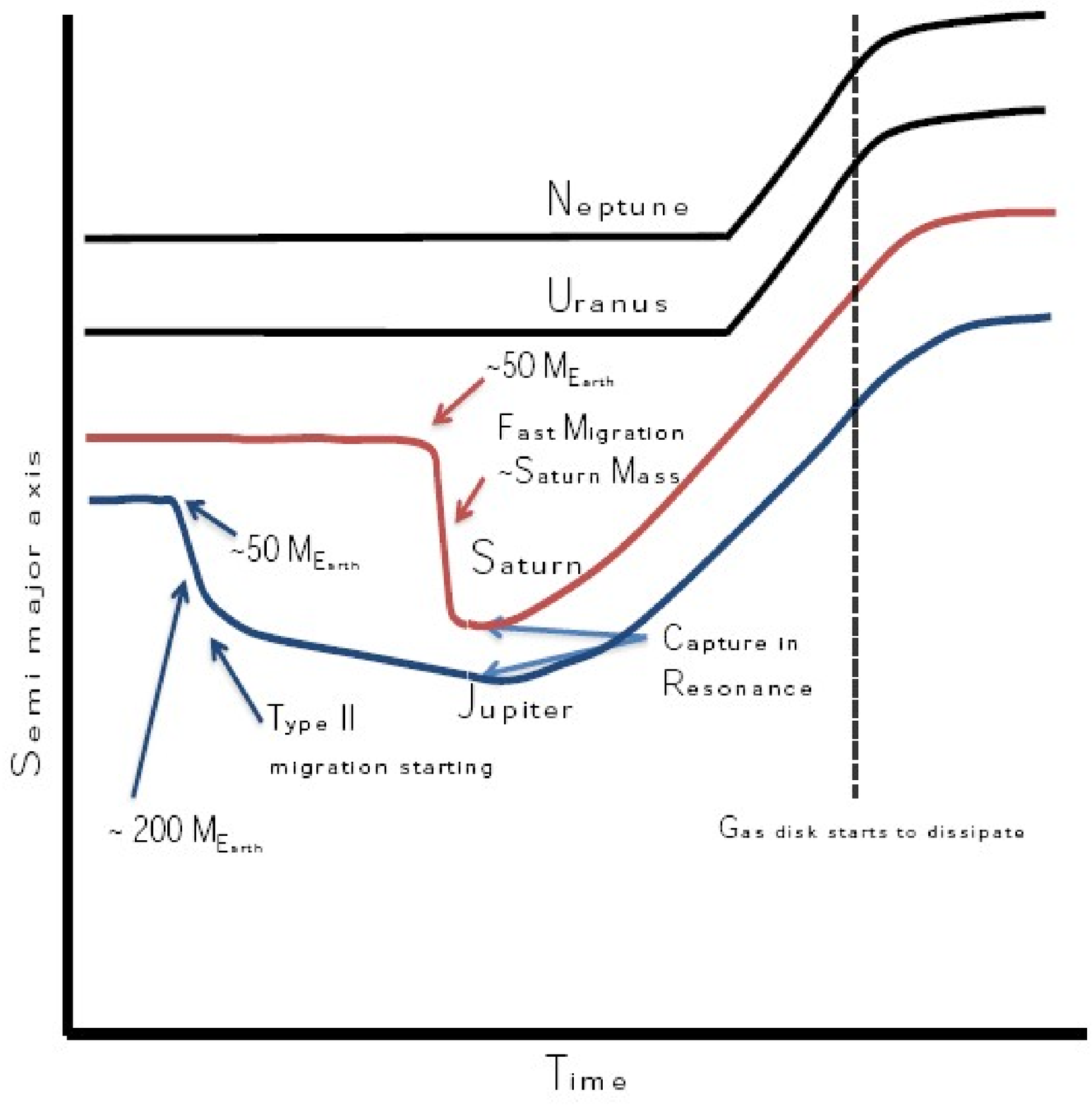}}
\vspace*{-.3cm} 
\caption{\small Sketch of the possible migration and accretion
 histories of the giant planets of the solar system, while they were
 embedded in the disk of gas. Both scenarios assume that: (i) Saturn
 was eventually caught Jupiter in their mutual 2/3 resonance (see
 sect.~\ref{TypeII} for conditions) and (ii) planets with masses
 smaller than $\sim 50$ Earth masses do not migrate in radiative
 disks. In the left panel, Jupiter and Saturn do not migrate after
 resonance capture. This leaves a large gap between the Jupiter/Saturn
 pair and the Uranus/Neptune pair, that cannot be reconciled (to our
 current knowledge) with the current orbital structure of the
 planets. In the right panel, Jupiter and Saturn migrate
 outwards. This leads to the capture of Uranus and Neptune in
 resonance with Saturn and with themselves. This resonant
 configuration is consistent with the current orbits of the planet via
 a phase of dynamical instability after the dispersal of the disk of
 gas (see sect.~\ref{inRes}).}
\label{Sketch} 
\end{figure}

What about Uranus and Neptune? According to the latest models on
migration of planetary cores in radiative disks, Uranus and Neptune,
or their precursors, should have evolved on non-migrating orbits in the
intermediate part of the disk (Lyra et al., 2010). 
If Uranus and Neptune did not migrate while Jupiter and Saturn
achieved non-migrating orbits after a period of inward migration,
eventually the giant planets system should have been
characterized by a large separation between the Jupiter/Saturn pair
and the Uranus/Neptune pair (see left panel of Fig.~\ref{Sketch}). How to
reconcile this peculiar orbital structure with the current structure of the
solar system is unknown. Instead, if Jupiter and Saturn migrated
outwards, Uranus and Neptune would have been caught in mean
motion resonances with the two major planets (see right panel of
Fig.~\ref{Sketch}). The same would have occurred also in the case
where Jupiter and Saturn remained on non-migrating orbits but Uranus
and Neptune started to migrate inwards due to a change of the thermal
properties of the disk over time (Lyra et al., 2010).  What matters,
in fact, is the convergent migration between the Jupiter/Saturn pair
and the Uranus/Neptune pair, not which pair of planets is moving.

A search for possible resonant configurations that could have been
achieved by the two pairs of planets in convergent migration has been
done in Morbidelli et al. (2007), using hydro-dynamical simulations,
and in Batygin and Brown (2010), using N-body simulations with
fictitious forces that mimic the effect of the disk.  The search is
probably not complete, in the sense that other relative resonant
configurations might have been achieved, depending on disk properties,
migration speed and initial Uranus/Neptune configuration.  Three
results, however, seem robust. First, in case of convergent migration,
each planet ends up in resonance with its neighbor.  Thus, the planets
should have been in a 4-body-resonance, the most complex
multi-resonant configuration in the solar system\footnote{Currently
the record for the most complex resonance chain is detained by
Jupiter's satellites Io, Europa and Ganymede, which are locked in a
3-body-resonance, also known as the Laplace resonance.}. Second,
because distant resonances are weak and have a low probability to
capture a migrating body, the four planets most likely should have
achieved a very compact orbital configuration. Third, the
eccentricities of the planets should have remained small (less than
0.005 for Jupiter, 0.02 for Saturn, 0.06 for Uranus and 0.015 for
Neptune in Morbidelli et al. simulations). Thus Jupiter and Saturn
should have been on orbits significantly more circular than now.

Morbidelli et al. also investigated the long-term stability of
resonant orbital configurations that they found. This was done by
continuing each hydro-dynamical simulation for 1,500 Jupiter's orbits,
while removing uniformly the gas, exponentially in time, down to a
factor of 1/1,000. This procedure was just instrumental for
changing adiabatically the potential felt by the planets, and was not
intended to mimic the real process of evaporation of the disk. The
final orbits of the planets were then passed to a symplectic N-body
code, and integrated for 1~Gy, without additional perturbations (for
instance from a planetesimal disk). Morbidelli et al. found that only
one configuration was stable for 1~Gy. This result was actually
affected by an error of the integrator, which was discovered only
later. In reality, 4 of the 6 configurations found in Morbidelli et
al. are stable for 1~Gy. The only
unstable configurations are the two most compact ones, with Uranus in
the 3/4 resonance with Saturn and Neptune in either the 4/5 or 5/6
resonances with Uranus. These configurations lead to very violent
instabilities, with close encounters of all the planets with each
other, including close encounters of Jupiter with Saturn. In these
cases, orbital relaxation is achieved when all planets except Jupiter
are ejected on hyperbolic or distant eccentric orbits.  Obviously this
is not what happened in our solar system. But, in these simulations,
the final orbit of Jupiter, with an eccentricity of 0.4, is similar to
those of the extra-solar planets discovered to date at a distance of
4--5 AU from their parent stars. This suggests that these extra-solar
planets might be the survivors of systems that avoided Type~II
migration, possibly through a mechanism like the Jupiter-Saturn one,
but which achieved orbital resonant configurations so compact to
undergo a violent instability involving encounters between gas-giant
planets. Our solar system was more lucky, and picked up a less compact
multi-resonant configuration, so that encounters between Jupiter and
Saturn could be avoided.

We will see in the next section that the least compact
resonant configurations, which are stable when only the 4 planets
(Jupiter to Neptune) are considered, can become unstable when the
interactions of the planets with a remnant planetesimal disk are taken
into account.  I will argue that the solar system must have passed
through such an instability phase, to reconcile the current orbits
with those that the planets should have had when they emerged from the
gas-disk phase (see also Thommes et al., 1999).

\section{The planetesimal-disk era}
\label{planetesimals}

\subsection{Brief tutorial of planetesimal-driven migration}
\label{tutor}

A planet embedded in a planetesimal disk has repeated close encounters
with the objects that come close to its orbit. Each of these
encounters modifies the trajectory of the incoming planetesimal and,
consequently, the planet has to suffer a small recoil.  In this way,
if the planetesimal disk is sufficiently massive, significant angular
momentum exchange may occur between the planet and the planetesimals,
enough to cause a rapid, long-range migration of the planet (Ida et
al., 2000). A review of planetesimal-driven migration has been
presented in Levison et al. (2007). In this section, I discuss the
basic concepts that are relevant to understand the evolution of our
solar system and, possibly, of planetary systems in general.

For a system of giant planets, planetesimal-driven migration is relevant
only after the disappearance of the gas. The reason is that the gas
contains typically $\sim 100$ times more mass than the planetesimals
and therefore it exerts the dominant gravitational forces on the
planets. Consequently, planetesimal-driven migration was not mentioned
in the previous part of this chapter. 
It should be remembered, though, that for small planets
gas-driven migration is proportional to the planet's mass (Ward,
1997), whereas planetesimal-driven migration is, at first order,
independent of the planet's mass (Ida et al., 2000; Kirsh et al.,
2009); thus, for small planets, such as planetary embryos of
an Earth-mass or smaller, planetesimal-driven migration may rival,
under some conditions, Type-I migration (Levison et al., 2010;
Capobianco et al., 2011). 

One may naively think that planetesimals scattering is a purely random
process, which consequently cannot force a planet to migrate in a
specific direction. This is not true. To zeroth order, during an
encounter the planetesimal is in a Keplerian orbit about the planet.
Since the energy of this orbit must be conserved, all the encounter
can do is to rotate the relative velocity vector.  Thus,
the consequences of such an encounter can be effectively computed in
most of the cases using an impulse approximation (Opik,~1976; Ida et
al., 2000). With this approach, it is easy to compute that on average
(that is averaged on all impact parameters and relative orientations)
the planetesimals that cause a planet to move outward are those whose
$z$-component of the specific angular momentum $H=\sqrt{a(1-e^2)}\cos{i}$ is
larger than that of the planet ($H_p$). In fact, during the encounter
these planetesimals have an azimuthal angular velocity faster than
that of the planet: thus, they tend to be slowed down, propelling in
turn the planet along its orbit.  The opposite is true for the
planetesimals with $H<H_p$ (Valsecchi and Manara,~1997).  In these
formul\ae\ $a$, $e$ and $i$ are the semi major axis, eccentricity and
inclination of the planetesimal. 

The direction of migration for a single planet in principle depends on
the angular momentum distribution of objects on planet-encountering
orbits. Nevertheless, the direction of migration does not simply
depend on the sign of $\bar{H}-H_p$, where $\bar{H}$ denotes the
mass-weighted value of $H$ of the
planet-crossing particles: there is a bias in scattering timescales on
either side of the planet's orbit which leads to a very strong
tendency for the planet to migrate inwards (Kirsh et al.,
2009). Consequently, outward migration is found only in systems where
$\bar{H}-H_p$ is strongly positive, such as for planetesimal disks
with surface density distribution proportional to $r^k$ with $k>1$
(the value of $k$ for the real disks is expected to be between $-2$
and $-1$, definitely giving inward migration).

\begin{figure}[t!]
\centerline{\psfig{figure=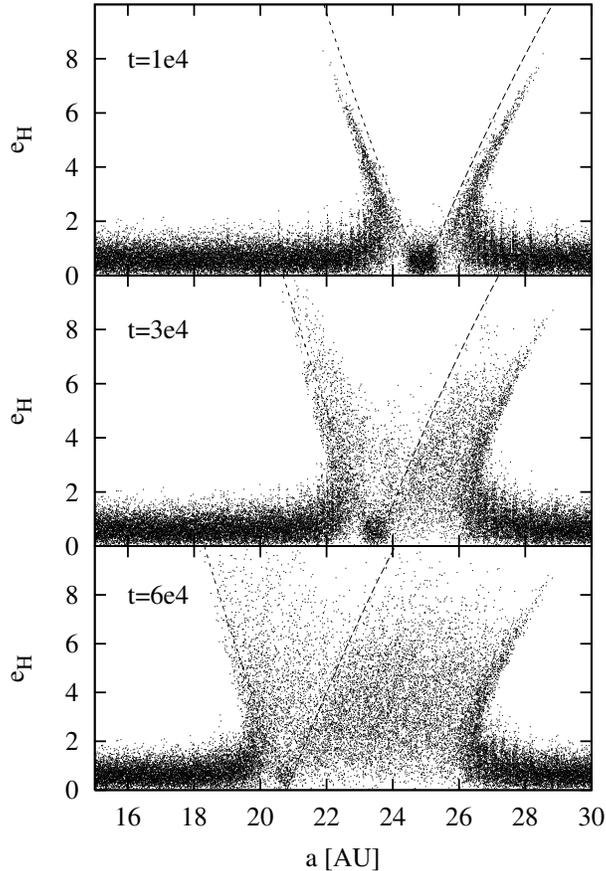,height=12cm,angle=-90}}
\vspace*{-.3cm} 
\caption{\small The migration of a 2.3~$M_\oplus$ planet in a
 planetesimal disk of 230~$M_\oplus$. Each panel shows the
 eccentricity vs. semi major axis distribution of the planetesimals
 (dots) at the time marked in the top left corner. The dashed curves
 delimit the planet crossing region. The planet is situated at the
 point of intersection of these curves. Notice how planetesimals are
 ``left behind'' on eccentric orbits as the planet migrates
 inwards. Adapted from Kirsh et al. (2009).  }
\label{Kirsh} 
\end{figure}

In order to understand the basic modes of migration, imagine defining
a function ${\cal F}$ of the $H$-distribution of the planet-crossing
particles, such that the planet's migration rate ${\rm d}a_p/dt$ is
proportional to ${\cal F}$ (which implies that migration is inwards if
${\cal F}$ is negative, while it is outwards otherwise).  For a given
orbital distribution of the planetesimals in the planet-encountering
zone (i.e. for a fixed value of ${\cal F}$), the rate of migration in
the local units of length and time must be proportional to the
total mass $M$ of the planet-encountering planetesimals (Ida et al.,
2000; Gomes et al., 2004; Levison et al., 2007). Thus, we can write
$$
{{{\rm d}a_p/a_p}\over{{\rm d}t/T}}\sim M{\cal F}\ , 
$$ 
where $T$ is the planet's orbital period.  When a planet migrates,
several concomitant processes occur. (a) The planetesimals that are
scattered in the direction opposite to that of planet migration
(e.g. onto an orbit with a larger semi major axis, for an inward
migrating planet) can be ``left behind'', in the sense that they can
find themselves on orbits which do not cross the planet's orbit any
more (because the orbit of the planet has drifted away; see
Fig.~\ref{Kirsh}). Conversely, other planetesimals, originally
situated on stable orbits in the disk through which the planet is
moving, start to be scattered. Thus, planetesimals both enter and
leave the planet-encountering region as a result of the radial drift
of the planet's orbit through the disk. This orbital drift tends to
leave ${\cal F}$ unchanged, but changes $M$; the latter increases or
decreases depending on the gradient of the disk's surface density. (b)
The planet encounters tend to re-arrange the angular momentum
distribution of the planetesimals to an equilibrium configuration that
would induce no-migration; this decreases $|{\cal F}|$ towards zero,
while preserving $M$. (c) Some planetesimals may be eliminated from the
system through collisions with the planet or ejections onto hyperbolic
orbit, which decreases $M$.  Depending on whether, as a net result of
all these processes, $|M{\cal F}|$ increases or decreases, radial
migration accelerates exponentially or decays to zero. The former case
is called {\it forced} or {\it self-sustained} migration; the latter
is called {\it damped migration}. Whereas processes (b) and (c)
inexorably tend to damp the planet's migration (because they reduce
either $|{\cal F}|$ or $M$), process (a) can sustain the migration if
it leads to an increase of $M$.  Thus self-sustained migration occurs
if two conditions are met: first, process (a) has to give a positive
feedback on migration, which translates into a condition on the
gradient of the surface density of the disk; second, the mass $M$ has
to be large enough so that the timescale of process (a), which is
proportional to $1/M$, is faster than those of (b) and (c), which are
independent of $M$.

The dynamical evolution is qualitatively different if there are two
planets. In this case, planetesimals can be scattered by one planet
onto an orbit that has close encounters with the other planet. This
transfer of particles from the ``control'' of one planet to the other
tends to increase the orbital separation between the planets.
However, the planets can effectively move away from each other only if
they are not locked in a mutual mean motion resonance (the orbital
response of resonant planets is different and will be discussed in
Sect.~\ref{inRes}). Assuming no resonance locking, under some
conditions, this orbital divergence can lead the outer planet to
migrate outwards, even in planetesimal disks in which a single planet
would normally migrate inwards. In particular, this is the case for a
Neptune-mass planet on an orbit exterior to a Jupiter-mass planet. In
fact, the planetesimals that the Neptune-mass planet scatters inwards
onto orbits crossing that of the Jupiter-mass planet, are rapidly
ejected onto hyperbolic orbits by the latter; conversely, the
planetesimals that the Neptune-mass planet scatters outwards, remain
on orbits crossing that same planet, and have repeated encounters with
it: sooner or later, most of them will be eventually scattered inwards
and then removed by an encounter with the Jupiter-mass planet. In
conclusions, the net work of the Neptune-mass planet is to transfer
planetesimals inwards to the control of the Jupiter-mass planet, and
consequently the Neptune-mass planet has to move outwards.

The case of the giant planets of our solar system, with Jupiter in the
innermost orbit, two Neptune/Uranus-mass planets on the outermost
orbits and one intermediate-mass planet (Saturn) in between, is
somewhat analog to the simple Jupiter-Neptune system described above
(if, again, the planets are not in resonance and are free to migrate
relative to each other). With $N$-body simulations, Fernandez and Ip
(1984) showed for the first time that Jupiter migrates inwards, while
Saturn and, particularly, Uranus and Neptune move outwards. Malhotra
(1993, 1995) elaborated on this kind of evolution to explain the
observed orbital properties of the Kuiper belt (a population of
planetesimals, including Pluto, with semi major axes beyond that of
Neptune). The original version of the ``Nice model''\footnote{Named
for the French city of Nice, where it was developed.}  (Tsiganis et
al., 2005; Gomes et al., 2005), that aimed to build a coherent
scenario of the late orbital evolution of the outer solar system, was
also based on this process of divergent migration of the giant
planets. Given our current understanding that the giant planets, at
the end of the gas-disk era, had to be in resonance with each other
(see previous section), these models are not strictly valid any more,
and consequently I won't discuss them further. A new version of the
``Nice model'', which starts from a multi-resonant orbital
configuration of the giant planets, is instead illustrated in
sects.~\ref{inRes} and~\ref{LHB}.

There are nevertheless some important results of general relevance
that these studies brought to light. The concepts of self-sustained
migration and damped migration apply to the multi-planet case as
well. In the case of self-sustained migration, Neptune tends to
migrate to the outer edge of the planetesimal disk, or at least up to
a large distance ($\sim 100$~AU) from the Sun (Gomes et al., 2004). In
the case of damped migration, the planets attain a final orbital
configuration after having removed all planetesimals in between them
and leaving a massive disk of objects a few AUs beyond the final orbit
of Neptune (Gomes et al., 2004). Given that the Kuiper belt contains
very little mass (probably less than 0.01~$M_\oplus$; Fuentes and
Holman, 2008) and excluding, from considerations based on its orbital
and size distributions, that said little mass is the result of collisional
grinding (Morbidelli et al., 2008b; see however Kenyon et al, 2008 for
an opposite view), my conclusion is that the original planetesimal
disk in our solar system had to have an effective outer edge at $\sim 30$~AU,
close to the current location of Neptune (Gomes et al., 2004;
Morbidelli et al., 2008b). 

Another general result is that, during planetesimal-driven migration,
the eccentricities of the planet are damped. This is related to a
process called dynamical friction, well known in models of planet
formation. In essence, dynamical friction is the mechanism by which
gravitating objects of different masses exchange energy so as to
evolve towards an equipartition of energy of relative motion (Saslaw,
1985): as a general rule, for a system of planets embedded in a
massive population of small bodies, the eccentricities and
inclinations of the former are damped, while those of the latter are
excited (Stewart \& Wetherill, 1988). Thus, during planetesimal-driven
migration, the only possibility for the enhancement of the
planets' eccentricities is that the planets pass through mutual
resonances as their orbits diverge from each other (Chiang, 2003;
Tsiganis et al., 2005).

\subsection{Multi-resonant planet configurations and planetesimal
  scattering: the Solar System case}
\label{inRes}

When two planets are in a mutual mean motion resonance, their orbits
cannot move freely relative to each other: the resonance holds
constant the ratio between the orbital periods, i.e. between the
orbital semi major axes. Thus, both planets have to migrate at the
same relative rate.  In this case, the differential forces that act on
the planets from the planetesimals disk modify instead the orbital
eccentricities of the planets.  More precisely, if the interaction
with the planetesimals is such that, in absence of the mean motion
resonance, the planets would suffer divergent migration (i.e. the
ratio between the semi major axes of the outer and the inner planets
would increase), the eccentricities of the planets decrease (Henrard,
1993). When the eccentricities are low enough, the planets can
eventually exit their resonance.  At this point,
if the planets are still stable, normal divergent migration, as
discussed in the previous section, can start.

\begin{figure}[t!]
\centerline{\includegraphics[height=6.cm]{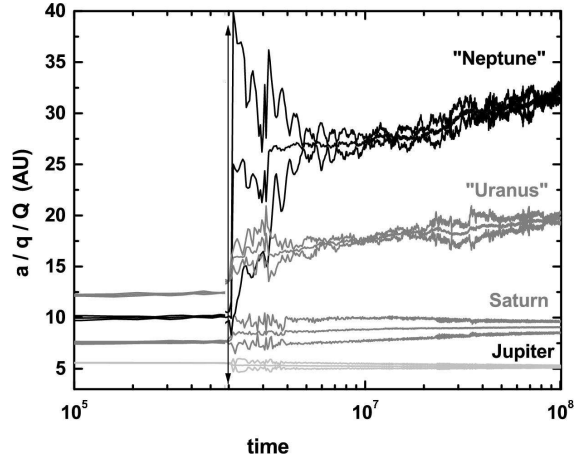}}
\vspace*{-.3cm} 
\caption{\small The evolution of the 4 giant planets of the solar
 system, starting from a 4-body resonance and embedded in a
 50~$M_\oplus$ planetesimal disk. Here, each planet is represented by
 three curves, showing the perihelion distance $q$, the semi major
 axis $a$ and the aphelion distance $Q$, respectively. Initially
 Saturn is in the 2/3 resonance with Jupiter, Uranus is in the 2/3
 resonance with Saturn and Neptune is in the 3/4 resonance with
 Uranus. The vertical arrow marks the time of the instability, when
 the orbits of the planets are extracted from the original 4-body
 resonance and become eccentric. In this simulation, Uranus and
 Neptune are scattered outwards by encounters with Saturn and between
 themselves. The final orbits are quite similar to the current
 orbits. From Morbidelli et al. (2007).}
\label{TsigII} 
\end{figure}

Let's consider now the case of the solar system. Assume for instance
that the planets were in the least compact of the multi-resonant
configurations found in Morbidelli et al. (2007): Saturn is in the 2/3
resonance with Jupiter, Uranus is in the 2/3 resonance with Saturn and
Neptune is in the 3/4 resonance with Uranus. Assume that the system,
at the disappearance of the gas, is still embedded in a planetesimal
disk, of about 50~$M_\oplus$, with an outer edge at $\sim
30$~AU. Fig.~\ref{TsigII} shows what happens: after a short time, the
planets are extracted from their multiple resonance. Resonances have a
strong stabilizing effect for close orbits (a clear example is that of
Pluto which, despite it crosses the orbit of Neptune, is stable
because it is in the 2/3 resonance with it). Once the planets are
extracted from their mutual resonances, this stabilizing effect
ends. The planets rapidly become unstable, because they are too close
to each other. In the simulation illustrated in
Fig.~\ref{TsigII}, the onset of the instability occurs at $t\sim
2$~My. Uranus and Neptune start to have close encounters with each
other and with Saturn. As a result, the orbits of all the planets become
more eccentric; Uranus and Neptune are propelled outwards by
encounters with Saturn, onto very eccentric orbits. Their orbital
eccentricities are then strongly damped by dynamical friction, which
stabilizes the motion of the planets and prevents further mutual close
encounters; the eccentricities of Jupiter and Saturn are also damped,
but to a lesser extent. At the same time, planetesimal-driven
migration, which can operate once the planets are extracted from
resonance, increases further the orbital separations between each pair
of planets. At the end of the simulation, the planets have acquired
orbits very similar to their current ones: the semi major axes are
within 10\% of the real values, and the eccentricities and
inclinations are also comparable. Most of the planetesimal disk has
been dispersed by then, so that little orbital evolution is expected
to occur after 100~My. 

This simulation shows that the multi-resonant configuration, which the
giant planets should have been driven into during the gas-disk phase,
is not incompatible with the current orbital configuration: the
interaction with the planetesimals disk and the temporary phase of
global instability, which the planets experience after extraction from
their original resonances, can very well lead the system to its
current dynamical state (Thommes et al., 1999). More examples of this
kind of successful evolution, starting also from different
multi-resonant orbital configurations, can be found in Batygin and
Brown (2010).

\subsection{The Late Heavy Bombardment as a smoking gun for a late
  instability of the giant planets}
\label{LHB}

In Fig.~\ref{TsigII} the dynamical instability occurs early, after
only 2~My from the beginning of the simulation. However, there is a strong
indication that in our solar system 
the onset of the dynamical instability happened much later,
approximately 600~My after the disappearance of the disk of gas: this
piece of evidence comes from the so-called ``Late Heavy Bombardment''
(LHB).  

The LHB is a cataclysmic period between $\sim 4.0$ and $\sim 3.8$~Gy
ago, marked by an extraordinarily high rate of collisions on the Moon
(Tera et al., 1974; Ryder, 1990, 2002; Cohen et al., 2000; Ryder et
al., 2000).  Some authors still contend the existence of such a
spike in the history of the bombardment rate (see for instance
Baldwin, 2006; Hartmann et al., 2007) and interpret the high
bombardment rate $\sim 3.9$~Gy ago as the tail of a slowly declining,
even-more-intense bombardment occurring since the time of formation of
the terrestrial planets. However, this seems to be implausible, for
several reasons:

\noindent{\it i)} 600 million years of continual impacts should have
left an obvious trace on the Moon.  So far, no such trace has been
found. The isotopic dating of the samples returned by the various
Apollo and Luna missions revealed no impact melt-rock older than
3.92~Gy (Ryder, 1990; Ryder et al. 2002). The lunar meteorites confirm this
age limit.  The meteorites provide a particularly strong argument
because they likely originated from random locations on the Moon
(Cohen et al., 2000), unlike the lunar samples collected directly on
its surface. A complete resetting of all ages all over the Moon is
possible (Hartmann et al., 2000) but highly unlikely, considering the
difficulties of completely resetting isotopic ages at the scale of a
full planet (Deutsch and Scharer, 1994).  The U-PB and Rb-Sr
isochrones of lunar highland samples indicate metamorphic
events between 3.85 and 4~Gy ago (Tera
et al., 1974).  There is no evidence for these isotopic systems being
reset by intense collisions between 4.4 and 3.9~Gy.

\noindent{\it ii)} The old upper crustal lithologies of the Moon do not 
show the expected enrichment in siderophile elements (in particular the
Platinum Group Elements) implied by a period of intense collisions
(Ryder et al., 2000) lasting 600~My.

\noindent{\it iii)} If the elevated mass accretion documented in the
period around 3.9~Gy is considered to be the tail end of an extended period of
even more intense collisions, the Moon should have reached 95\% of its
total mass about 4.1~Gy ago instead of
4.5~Gy ago (Ryder, 2002; Koeberl, 2004).  

\noindent{\it iv)} Given the fast dynamical and collisional decay of
the population of planetesimals that remain in the vicinity of the
Earth's orbit at the end of the accretion process of the 
terrestrial planets, the formation of two huge
impact structures such as the Imbrium and Orientale basins (and
probably many more) on the Moon 600~My later implies an implausible
initial total mass of solids in the inner solar system (Bottke et al., 2007).

\noindent{\it v)} The bombardment rate 3.8-3.9 Gy ago (as deduced from
the lunar crater record) was probably not intense enough to vaporize
the oceans on Earth (Abramov and Mojzsis, 2009). However, if this
bombardment rate had been the tail of a more intense bombardment,
smoothly decaying over time since lunar formation, the ocean
evaporation threshold should have been overcome just a few hundreds of
millions of years earlier ($\sim$ 4.2 Gy ago). This contrasts with the
oxygen isotopic signature of the oldest known zircons (age: 4.4~Gy),
which indicates formation temperatures compatible with the existence
of liquid water (Valley et al., 2002).

\noindent{\it vi)} These same zircons retain secondary over-growths
developed after primary core crystallization during their 4.4~Gy long
crustal residence times. The rim over-growths can record discrete
thermal events subsequent to zircon formation and provide a unique
window in crustal processes before the beginning of the terrestrial
rock record. In (Trail et al., 2007), all these rim over-growths have
been dated to be $\sim 3.9$~Gy old. No (preserved) older rim
over-growths, associated to more primordial events, have been
found. This suggests that the thermal events were associated to
impacts, and that these impacts were concentrated in time about 3.9~Gy
ago.

Therefore, it can be concluded that there is strong evidence for a cataclysmic
Late Heavy Bombardment event around 3.9~Gy ago.  This cataclysm did not
just affect the Moon, but has now been clearly established throughout the
inner Solar System (Kring and Cohen, 2002). The exact duration of the cataclysm
is difficult to estimate, however.  Based on the cratering record of the Moon,
it lasted between 20 and 200~My, depending on the mass flux estimate used in
the calculation.

The very existence of an LHB implies that a massive population of
planetesimals must have been stored for $\sim 600$~My in a stable
reservoir, which then was suddenly destabilized. The only intuitive
way to do this is that there was a sudden change in the orbital
structure of the planets at that time (Levison et al., 2001). In fact,
the planets, in particular the giant ones, control the dynamical
evolution of the small bodies and determine which reservoir is stable
and which unstable. The planetary evolution illustrated in
Fig.~\ref{TsigII} would do a great job in causing an impact spike, by
destabilizing the full outer planetesimal disk when Uranus and Neptune
are propelled to their current orbital semi major axes; the asteroid
belt would also be partially destabilized when Jupiter and Saturn
acquire their current eccentricities and move towards their current
orbital separation (Gomes et al., 2005; Minton and Malhotra, 2009).
However, in the simulation of Fig.~\ref{TsigII} the spike would occur
too early to coincide with the LHB spike: we need to find a plausible
explanation for which the extraction of the giant planets from their
original multiple resonance occurred not after only 2~My (as in
Fig.~\ref{TsigII}), but approximately 600~My later.

The reason for which, in the simulation of Fig.~\ref{TsigII} (and in
the simulations of Batygin and Brown, 2010), the
instability occurs early is that the planets were assumed to be {\it
embedded} in a planetesimal disk, so that the interaction with said
disk was very strong. As pointed out in Gomes et al. (2005), however,
this is an unlikely configuration. In fact, the planetesimals that are
originally in between the orbits of the giant planets are violently
unstable, with a dynamical lifetime well shorter than 1~My. Thus, they
should have been removed (by colliding with the planets, being ejected
onto distant orbits etc.) well before the disappearance of the gas,
which typically lasts 3--5 My in a protoplanetary disk (Haisch et al.,
2001). Then, as said at the beginning of sect.~\ref{planetesimals},
the early removal of these planetesimals should not have changed the
orbital configuration of the planets, because the dominant forces
exerted by the disk of gas forced the planets to stay in their multiple
resonance. Therefore, it is more likely that, at the disappearance of
the gas, when $N$-body simulations like that of Fig.~\ref{TsigII}
become relevant, the planetesimals were only on those orbits whose
dynamical lifetime is of the order of the gas-disk lifetime, or
longer. This constrains the planetesimals to be in a trans-Neptunian
disk, with an inner edge situated at least 1--2~AUs beyond the original semi
major axis of Neptune (for simplicity, I call here ``Neptune'' the
planet that is the most distant from the Sun; notice that in some
simulations -that of Fig.~\ref{TsigII} for instance- the two last
planets in order of distance from the Sun switch orbits; in these
cases Uranus would have been originally the most distant planet from
the Sun).

\begin{figure}[t!]
\centerline{\includegraphics[height=7.cm]{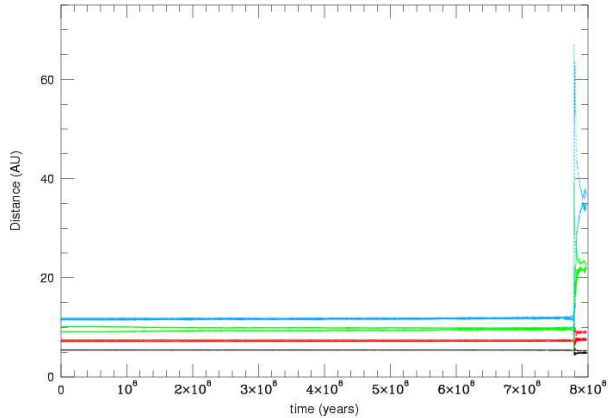}}
\vspace*{-.3cm} 
\caption{\small The same as Fig.~\ref{TsigII}, but for a planetesimal
  disk with an inner edge suitably placed beyond the initial orbit of
  the outermost planet. Here, the instability is delayed to $\sim
  700$~My, consistent with the timing of the Late Heavy Bombardment of
  the terrestrial planets.}
\label{figLHB} 
\end{figure}

If the planetesimal disk is beyond the orbit of Neptune, the
interactions between the planets and the disk are necessarily much
weaker than in the case where the planets are embedded in the disk. 
In this condition, the instability can occur late, after a time
comparable with the LHB chronology, as shown in Fig.~\ref{figLHB}. I
acknowledge, though, that the instability time depends critically on
the location of the inner edge of the disk: disks with inner edges
slightly closer to Neptune lead to early instabilities, and disks with
edges just a bit further give systems that are stable
forever\footnote{The situation was not nearly as sensitive in 
  Gomes et al. (2005), because the planets were not
  assumed to be in resonance with each other.}. Such extreme
sensitivity looks problematic.

All the simulations presented up to this point, however, were simple,
because they assumed that the planetesimals do not to interact
dynamically with each other. If self-interactions are taken into
account, for instance assuming that there are a few 100s Pluto-mass
objects in the disk perturbing each other and the other particles,
then there is a net exchange of energy between the planets and the
disk, even if there are no close encounters between planets and
planetesimals. This is because the self-stirring of the disk breaks
the reversibility of the eccentricity coupling between planet and
planetesimals: the planet eccentricities are damped; however, the
evolutions of orbital energy and eccentricity are coupled at second
order in the masses (Milani et al., 1987): this produces a drift in
the planet's energy.  In particular, the planets loose energy,
i.e. they try to migrate towards the Sun (Levison et al., 2011).  The
orbits of the planets tend to {\it approach} each other. This is
different from the case where planets scatter planetesimals, in which
the planetary orbits tend to {\it separate} from each other. Remember,
though, that the planets are in resonances; so the ratios between
their semi major axes cannot change. In response, the planetary
eccentricities slowly {\it increase}. This eventually drives some
planets to pass through secondary or secular resonances, which
destabilize the original multi-resonant configuration. The overall
evolution is very similar to what is presented in Fig.~\ref{figLHB}
but now the instability time is late in general: in the simulations of
Levison et al. (2011) it ranges from 350~My to over 1~Gy for disks
with inner edge ranging from 15.5 to 20~AU (Neptune is at $\sim
11.5$~AU in these simulations). Unlike the case without
self-interactions of disk particles, there is no apparent correlation
between instability time and initial location of the inner edge of the
disk. This may appear surprising, because the rate of energy exchange
between planets and disk decreases with increasing distance of the
disk's inner edge. However this dependence is weak, because the
planet-disk interaction is a distant interaction (no close encounters
are involved). Then, the expected monotonic dependence of the
instability time on the disk's distance can be easily erased by the
fact that the evolutions of the disk and of the planets are chaotic,
which gives a highly sensitive and non-trivial dependence of the
results on the initial conditions. The instability time seems to
depend weakly also on the number of Pluto-mass scatterers in the disk,
provided that this number exceeds a few hundreds.

Together, the papers by Morbidelli et al. (2007) and Levison et
al. (2011) build the new version of the ``Nice model''. This is much
superior than its original version (Tsiganis et al.,
2005; Gomes et al., 2005) because (i) it removes the arbitrary
character of the initial conditions of the planets by adopting as
initial configuration one of the end-states of hydro-dynamical
simulations and (ii) it removes the sensitive dependence of the
instability time on the location of the inner edge of the disk;
instead, a late instability seems to be a generic outcome.

In the new Nice model, only 10\% of the simulations which exhibit a
global dynamical instability of the planets lead to a
stable 4-planets system at the end; in the remaining simulations one
or more planets are lost, ejected onto hyperbolic orbits. The
fraction of ``successful'' simulations in the original simulations of
Tsiganis et al. (2005) was much higher: $\sim 50$\%. This is because
in the new model the planets are initially in a more compact
configuration and therefore their orbital instability is more
violent. This low probability of success may suggest that we are still
missing something important in our reconstruction of the past history
of the solar system, but on the other hand it is not so low to reject
the model a priori. After all, we have observed only one such system
so far! On the positive side, like in Tsiganis et al., when 4 planets
survive in the new Nice model, their final orbits are very similar to
the real ones: the orbital semi major axes are within 10-15\% of the
real values, and the final orbital eccentricities and inclinations are
also close (within a factor of 2) to the actual ones. This argues that
the model, although certainly not perfect, is probably not too far
from reality.

\subsection{The solar system as a debris disk: are LHBs common?} 

If our understanding of the evolution of the solar system is (even
approximately) correct, there should have been a massive belt of
planetesimals outside Neptune's orbit during the first $\sim
600$~My of history, i.e. up to the LHB time.  This disk, through
mutual collisions, should have produced a large amount of dust,
generating what it is usually referred to as a ``debris disk''. It is
interesting to investigate how our debris disk would have appeared to
an extra-solar observer, and compare the result to the debris disks
that we infer around other stars of various ages.

Booth et al. (2009) addressed this question. The intrinsic collision
probabilities and mutual velocities among the planetesimals have been
computed from the dynamical simulations of the Nice model (in its old
version, but this should not make a big difference in this
respect). To compute the outcome of the collisional activity, Booth et
al. had to assume an initial size distribution. They adopted the size
distribution observed in the current Kuiper belt, with the number of
objects in each size-bin multiplied by a factor $\sim 1,000$. Thus the
initial size distribution contained $\sim 1,000$ Pluto-size bodies,
was relatively steep down to $D\sim 100$~km, and then turned over to a
shallower slope for sizes smaller than this threshold (Bernstein et
al., 2004; Fuentes and Holman, 2008). With this assumption, a
planetesimal disk with initially 50~$M_\oplus$ of material looses less
than 50\% of its mass in 600~My because of collisional grinding (see
also the supplementary material of Levison et al., 2009). This is an
important consistency check for the Nice model: in fact, if collisional
grinding rapidly removed most of the mass of the disk no matter the
initial size distribution, then there would not be enough mass to
affect the evolution of the giant planets at the LHB time.

\begin{figure}[t!]
\centerline{\includegraphics[height=10.cm]{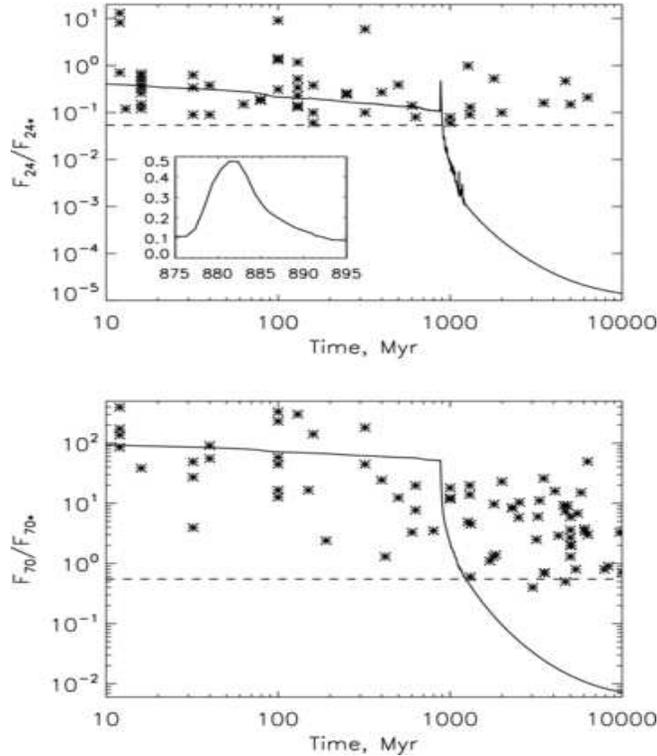}}
\vspace*{-.3cm} 
\caption{\small The infra-red luminosity of the debris disk of our
  solar system, according to the Nice model. The solid line shows the
  luminosity of the disk relative to that of the Sun at 24 microns
  (top) and 70 microns (bottom). The luminosity decays slowly during
  the first 880~My, when the planets become unstable (in the adopted
  simulation). Then the luminosity of the disk rapidly decays, as the
  planetesimals are removed from the solar system. The horizontal
  dashed line shows the detection limit for an extra-solar
  observer with our current measurement capabilities. The asterisks
  represent the observed disks. The window in the lower left corner of
  the top panel is a magnification of the evolution around the
  instability time. From Booth et al. (2009).}
\label{Booth} 
\end{figure}

After having computed the dust production rate as a function of time
and the orbital and size distributions of the grains, Booth et
al. computed the emission of the disk at wavelengths of 24 and 70
microns, relative to the emission of the Sun. The result is
illustrated in Fig.~\ref{Booth}. It shows that the luminosity of the
debris disk should have decayed very slowly, by less than an order of
magnitude, during the time preceding the onset of the planetary
instability (which occurred at 880~My in the specific simulation of
the Nice model used in their calculation). Then, the luminosity should 
have decayed very rapidly below detectability, as the disk was
dynamically dispersed by the planets. The figure also shows, with star
symbols, the infrared excess of known stars, as a function of their
estimated ages. At 24 microns, only $\sim 15$\% of the stars younger
than 300~My have this kind of excess (Carpenter et al., 2009), and
this fraction decreases to a few percent for older stars (Gaspar et
al., 2009); at 70 microns, the fraction of stars with detectable
infrared excess does not seem to decay with age (Hillebrand et al.,
2008).  

A first important conclusion, from the comparison of the estimated
brightness of our solar system with the IR-excess of other stars, is
that, before the LHB, our debris disk was fairly typical.  A second
conclusion, from the fact that there is no general tendency for a
sudden disappearance of the IR-excess at 70 microns around other stars
at about 1~Gy of age, is that a complete dynamical clearing of
the planetesimal disk like the one that occurred in our solar system at the
LHB time is fairly atypical. From a statistical analysis of the
data, Booth et al. estimated that at most 15\% of the extra-solar
systems undergo such a late dynamical clearing.

The fact that a {\it late} complete dynamical clearing of the
planetesimal disk is a rare event should not be a big surprise. In
fact, it is clear from what has been said above that the events
described in the Nice model depend on two specific properties that not
many planetary systems might have in common with our own. First, the
planets in our system did not migrate permanently into the inner solar
system; instead, they remained or returned near their birth places,
i.e. adjacent to the planetesimal disk that generated them. Thus, when
their orbits changed at the time of the instability, they strongly
affected the disk. If the planets had migrated close to the Sun and
had remained there, they
would have lost contact with the distant planetesimal disk. Even if
the planets had become unstable in the inner solar system, probably
the distant disk would not have been dynamically cleared. Second, the
planetesimal disk of the solar system was small, presumably truncated
at $\sim 30$~AU (Gomes et al., 2004). If the planetesimal disk had
been extended to much larger distances, the dynamical instability and
the migration of Neptune would have probably cleared the disk up to
50--100~AU; beyond this threshold the disk would have remained massive
and would have continued to produce dust.

Coming back to Fig.~\ref{Booth}, the little spike in the disk's
brightness visible at 880~My in the upper plot (magnified in the box
in the bottom left corner) is the signature of the LHB event, due to a
burst in collisional activity that occurs as the disk starts to be
dispersed and its orbital excitation suddenly increases. As one can
see, the spike is not prominent enough to make the disk stand out of
the natural distribution of brightness of disks of different masses
(suggested by the dispersion of the observations reported on the top
panel of Fig.~\ref{Booth}).  The Booth et al. calculation, however,
does not account for the huge flux of comets into the inner solar
system that should have occurred during the disk dispersal: these
comets should have liberated a great amount of dust once inside a few
AUs from the Sun. Nesvorny et al. (2010) accounted for this effect:
they estimated that the inner zodiacal cloud should have been more
than $10^4$ times brighter during the LHB epoch.  As the current
infrared excess at 24 microns of the zodiacal cloud is $2\times
10^{-4}$ (Kelsall et al., 1998), the excess at the LHB time at this
wavelength was probably of order 2--10. This kind of excess is
comparable to the upper envelope of the observations. Thus, the
conclusion is that luminosity bursts associated to LHB events, totally
invisible at 70 microns (cold dust), start to be detectable at 24
microns (hot dust), although they can still be confused with the
luminosity of massive disks undergoing gradual collisional
grinding. Therefore, the identification of systems that might be
undergoing an LHB event at the current time requires a case by case
analysis at multiple wavelengths, as done for instance in Wyatt et
al. (2007).

\section{Terrestrial planets}

Up to this point, this chapter was focussed on the formation and
evolution of giant planets. I now extend briefly the discussion to the
case of the terrestrial planets.

Terrestrial planets are expected to accrete in the inner part of a
proto-planetary disk, closer to the star than the snowline
position. Because of the absence of ices and of the relatively small mass
contained in the inner portion of the disk, the largest objects formed
by the processes of runaway and oligarchic growth (see
Sect.~\ref{form}) are expected to have masses of about a Lunar to a
Martian mass (Weidenshilling et al., 1997). These bodies are called
{\it planetary embryos} to distinguish them from the much more massive
{\it planetary cores} that are the precursors of the giant planets. 
They incorporate about 50\% of the mass of solids in the inner disk,
the rest remaining in asteroid-sized planetesimals. 

\begin{figure}[t!]
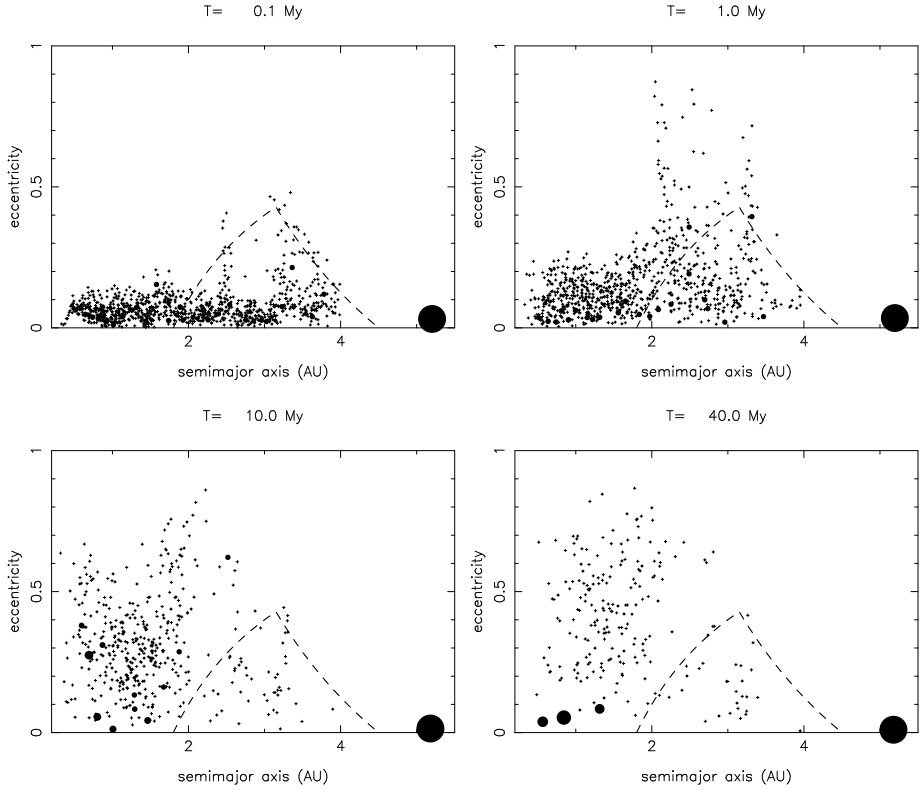

\centerline{\includegraphics[height=5.cm]{TPFae0002.ps}\, 
\includegraphics[height=5.cm]{TPFae0011.ps}}
\vskip 10pt
\centerline{\includegraphics[height=5.cm]{TPFae0101.ps}\, 
\includegraphics[height=5.cm]{TPFae0401.ps}}
\vspace*{-.3cm} 
\caption{\small Snapshots of the terrestrial planets formation
  process, from O'Brien et al., (2006). Each panel shows the
  eccentricity vs. semi major axis distribution of planetary embryos
  (filled circles) and planetesimals (crosses) at the reported
  time. The size of each filled circle is proportional to the cubic
  root of the mass of the corresponding embryo. The big filled ball at
  $\sim 5.2$ AU represents Jupiter. The dashed curves show the current
  boundaries of the asteroid belt. Notice the formation of three
  terrestrial planets in this simulation, the biggest of which is
  approximately one Earth mass. In the asteroid belt no terrestrial
  planet is formed. All embryos have left the asteroid belt region and
  only a small fraction of the initial planetesimals reside there on
  excited orbits.  In these simulations all collisions are supposed to
  be accretional. This approximation has been recently removed by
  Kokubo and Genda (2010), who considered a database of 
  collisions simulated by the SPH method (like in Agnor
  and Asphaug, 2004; Asphaug et al., 2006) to determine how much mass is
  accreted or ejected in each collsional event. The final results,
  though, show very little differences with respect to the simulations
  that treat all collisions as 100\% accretional.}
\label{TPs} 
\end{figure}

The evolution of the system of embryos and planetesimals has been the
object of many simulations using $N$-body integrations (Chambers and
Wetherill, 1998; Agnor et al., 1999; Chambers, 2001; Raymond et al.,
2004, 2005, 2006, 2007, 2009b; O'Brien et al., 2006; Kenyon and
Bromley, 2006; Kokubo et al., 2006; Thommes et al., 2008; Morishima et
al., 2010; Kokubo and Genda, 2010). All studies agree on the basic
aspects of the terrestrial planet accretion process, although they may
differ in the details. At the disappearance of the gas, the system of
embryos becomes violently unstable, due to the mutual interactions
among the embryos themselves and to the ``external'' perturbations
from the giant planets.  The orbits of the embryos begin to intersect
and accretional collisions between embryos start to occur. This
produces a smaller number of more massive objects (see
Fig.~\ref{TPs}), which are eventually identified with the terrestrial
planets.   

Concerning the planetesimals, a fraction of them contributes to the
growth of the planets by colliding with the embryos. The majority of
the planetesimals, however, are dispersed onto orbits with large
eccentricities and inclinations. In this process (the ``dynamical
friction'' mentioned in Sect.~\ref{tutor}), they damp the
eccentricities and inclinations of the growing planets, which
eventually sets the system into a stable configuration, with the most
massive planets on the least eccentric orbits.

This scenario for the formation of the terrestrial planets of our
solar system has several strong points:
\begin{itemize}

\item A system of 2--5 planets is typically formed. The efficiency of
  the accretion process is about 50\%. Thus, starting with $\sim 5 M_\oplus$
  in embryos and planetesimals 
  typically produces a couple of planets of about an Earth mass each
  (Chambers, 2001). The final orbits of the terrestrial planets
  produced in the simulations are comparable to those of the real
  terrestrial planets of our solar system, if the dynamical friction
  process is properly taken into account (O'Brien et al., 2006).

\item Quasi-tangent collisions of Mars-mass embryos onto the proto-planets are
  quite frequent (Agnor et al., 1999). These collisions are expected to
  generate a disk of ejecta around the proto-planets (Canup and Asphaug, 2001),
  from which a satellite is likely to accrete (Canup and Esposito, 1996). 
  This is the standard, generally
  accepted scenario for the formation of the Moon.
  
\item The accretion timescale of the Earth analog in the
  simulations is 30--100~My. This is in the good ballpark with the
  chronology of Earth accretion as indicated by
  radioactive chronometers, which still has a comparable uncertainty
  (Kleine et al., 2009).

\item In many/most simulations, terrestrial planets do not form in the
  asteroid belt. Instead, all the embryos are removed by mutual
  interactions and perturbations from Jupiter. A small fraction (a few
  percent) of the planetesimal population is left behind on stable
  asteroid-belt orbits, with eccentricities and inclinations
  comparable to those of the real asteroids (Petit et al., 2001;
  O'Brien et al., 2007). 

\item A significant fraction ($\sim 10$--20\%) of  the mass of the
  terrestrial planets is accreted from the outer part of the asteroid
  belt, which provides a
 formidable mechanism to explain the delivery of water to the Earth
 (Morbidelli et al., 2000; Raymond et al., 2004, 2007).

\end{itemize}

On the other hand, this scenario has a major problem: the planet
formed in the simulations at the location of Mars is typically too
massive (Chambers, 2001; Raymond et al., 2009; Hansen, 2009; Morishima
et al., 2010). Mars is an oddity not only for what it concerns its
mass, but also its accretion timescale: in fact, it formed in a few
millions of years only, like asteroids, i.e. much faster than the Earth 
(Dauphas and Pourmand, 2011). 
 
There does not seem to be a simple solution to the Mars
problem (Raymond et al., 2009). Hansen (2009) argued convincingly that
a correct mass distribution of the terrestrial planets, with an 
Earth/Mars mass ratio of $\sim 10$, can be achieved 
only if the initial disk of embryos and planetesimals is assumed to
have an outer edge at about 1 AU. The problem is how to justify such
an edge and how to reconcile this with the evidence that asteroids
exist in the 2--4 AU range. The simulations that assume Jupiter and
Saturn initially on orbits with their current separation in semi major
axis but eccentricities 2--3 times larger, do produce
very rapidly an effective edge at $\sim 1.5$~AU in the distribution of
embryos and planetesimals (Raymond et al., 2009; Morishima et al.,
2010) and result in a somewhat small ``Mars''. However, this initial
configuration of the giant planets is inconsistent with our
understanding of their orbital evolution through the history of the
solar system, described above (see Sects.~\ref{4plSS},~\ref{inRes}
and~\ref{LHB}). Moreover, none of these simulations are without
problems: Mars is often not small enough, the final distribution of
bodies in the asteroid belt is not good, water is not delivered to the
terrestrial planets etc..

\subsection{Linking giant planet migration to terrestrial planet
  accretion: the Grand Tack scenario}
\label{GT}

The result by Hansen motivated Walsh et al. (2011) to look in more
details at the possible orbital history of the giant planets and their
ability to sculpt the disk in the inner solar system. For the first
time, the giant planets were not assumed to be on static orbits (even
if different from the current ones); instead Walsh et al. studied the
co-evolution of the orbits of the giant planets and of the precursors
of the terrestrial planets, during the era of the disk of gas. 

\begin{figure}[t!]
\centerline{\includegraphics[height=10.cm]{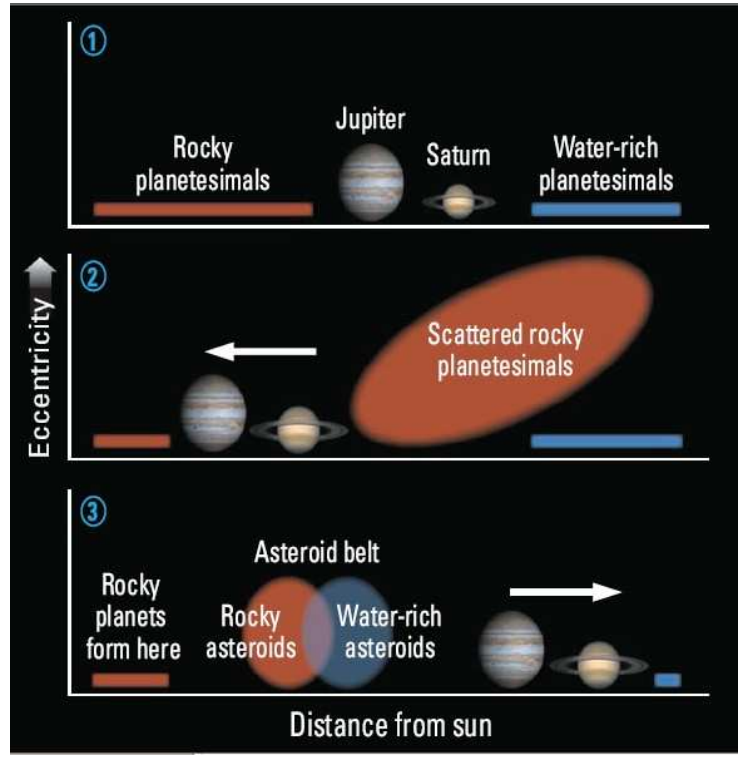}}
\vspace*{-.3cm} 
\caption{\small Sketch of the ``Grand Tack scenario''. The three
  panels depict subsequent steps in the evolution of the system. Form 
  ``NEWS\&ANALYSIS'': Science, {\bf 332}, 1255 (2011).}
\label{GTsketch} 
\end{figure}

Walsh et al. envisioned the following scenario, based on the considerations
reported in Sect.~\ref{TypeII}: first, Jupiter migrated inwards while
Saturn was still growing; then, when Saturn reached a mass close to
its current one, it started to migrate inwards more rapidly than
Jupiter, until it captured the letter in the 2/3 resonance; finally
the two planets migrated outwards until the complete disappearance of
the disk of gas. The extent of the inward and outward migrations
cannot be estimated a priori, because they depend on properties of the
disk and of giant planet accretion that are unknown, such as: the
time-lag between Jupiter and Saturn formation, the speed of inward
migration (depending on disk's viscosity), the speed of outward
migration (depending on disk's scale height), the time-lag between the
capture in resonance of Jupiter and Saturn and the photo-evaporation
of the gas. However, the extent of the inward and outward migrations
of Jupiter can be deduced by looking at the resulting structure of the
inner solar system. In particular, Walsh et al. remarked that a
reversal of Jupiter's migration at 1.5 AU would provide a natural
explanation for the existence of the outer edge at 1 AU of the inner
disk of embryos and planetesimals, required to produce a small Mars
(see Fig.~\ref{GTsketch}). Because of the prominent
inward-then-outward migration of Jupiter that it assumes, Walsh et
al. scenario is nicknamed ``Grand Tack''.

Several giant extra-solar planets have been discovered orbiting their
star at a distance of 1-2 AU, so the idea that Jupiter was sometime in
the past at 1.5 AU from the Sun is not shocking by itself. A crucial
diagnostic of this scenario, though, is the survival of the asteroid
belt. Given that Jupiter should have migrated through the asteroid
belt region twice, first inwards, then outwards, one could expect that
the asteroid belt should now be totally empty.  However, the numerical
simulations by Walsh et al. show that the asteroid belt is first fully
depleted by the passage of the giant planets, but then, while Jupiter
leaves the region for the last time, it is re-populated by a small
fraction of the planetesimals scattered by the giant planets during
their migration. In particular, the inner asteroid belt is dominantly
re-populated by planetesimals that were originally inside the orbit on
which Jupiter formed, while the outer part of the asteroid belt is
dominantly re-populated by planetesimals originally in between and
beyond the orbits of the giant planets (see
Fig.~\ref{GTsketch}). 

Assuming that Jupiter accreted at the location of the snow line, it is
then tempting to identify the planetesimals originally closer to the
Sun with the un-hydrous asteroids of S-type and those originally in
between and beyond the orbits of the giant planets with the 
``primitive'' C-type asteroids. With this assumption, the Grand Tack
scenario explains the physical structure of the asteroid belt probably
better than any other previous model. In fact, the asteroid belt is
characterized by a radial gradient in asteroid spectroscopic types
(Gradie and Tedesco, 1982): the inner belt is dominated S-type
(usually considered to be the parent bodies of ordinary chondrites;
Binzel et al., 1996), the outer belt by C-type asteroids (usually
considered to be the parent bodies of carbonaceous chondrites;
Burbine, 2000), although there is a significant overlapping between
the distributions of these different types of asteroids.  It is
difficult to explain the differences between ordinary chondrite and
carbonaceous chondrite parent bodies if they had both formed in the
asteroid belt region, given that they are coeval (Villeneuve et al.,
2009) and that the radial extent of the asteroid belt is small (~1 AU
only). Instead, if ordinary and carbonaceous chondrite
parent bodies have been implanted into the asteroid belt from
originally well separated reservoirs, as in the Grand Tack scenario,
the differences in physical properties are easier to understand in the
framework of the classical condensation sequence. The origin of C-type
asteroids from the giant planet region would also explain, in a
natural way, the similarities with comets that are emerging from
recent observational results and sample analysis (see sect~7 of the
supplementary material of Walsh et al., 2011, for a in-depth
discussion). The small mass of the asteroid belt and its eccentricity
and inclination distribution are also well reproduced by the Grand
Tack scenario.

This scenario also explains why the accretion timescales of Mars and
the asteroids are comparable (Dauphas and Pourmand, 2011). In fact,
the asteroids stopped accreting when they got dispersed and
injected onto excited orbits of the main belt; Mars stopped accreting
when the inner disk was truncated at 1 AU and the planet was pushed
beyond this edge by an encounter with the proto-Earth (Hansen,
2009). In the Grand Tack scenario these two events coincide, and mark
the time of the passage of Jupiter through the inner solar system. 

All these results make the Grand Tack scenario an appealing
comprehensive model of terrestrial planet formation and argue
strongly in favor of an evolution of the giant planets of our solar
system like that sketched in the right panel of Fig.~\ref{Sketch}.

\subsection{Terrestrial planets in extra-solar systems} 

Given that the architecture of the giant planets of our solar system
is far from being typical around other stars, it is interesting to
investigate the dependence of the terrestrial planet accretion process
on the properties of giant planets, across a wide
range of parameters. 

From various simulations (Levison and Agnor, 2003; Raymond et al.,
2004; 2006), it turns out that the outcome of the terrestrial planet
formation process has a weak dependence on the mass of the giant
planets. Obviously, the terrestrial planets cannot form in the vicinity
of giants. So, if the giant planets are closer to the star than
Jupiter, they leave to the terrestrial planets a narrower niche to
form inside. Instead, the process of terrestrial planets accretion is
very sensitive on the eccentricities of the giant planets. Large 
eccentricities of the giant planets force large eccentricities on embryos and
planetesimals. As a result, the final
terrestrial planets will be more eccentric; consequently they 
will have a
larger separation in semi major axis, will be less numerous and more
massive compared to a simulation where the same giant planets are on
circular orbits. Moreover, the planetesimals originally in the
vicinity of the giant planets (presumably rich in water and other
volatiles) are more likely to be ejected from the system than to
collide with the terrestrial planets, if the giant planets are
eccentric (Chambers and Cassen, 2002; Raymond et al., 2004). So, the
resulting terrestrial planets are expected to be water-poor.

The works quoted above assumed giant planets on ``fixed'' orbits. We
know now that the giant planets can have evolutions that
lead them to change their orbits, through
migration and/or dynamical instabilities. It is interesting to explore
how the terrestrial planets, during and after their formation, respond
to these changes. 

The effect of a Jupiter-mass planet migrating through the disk towards a
``hot-Jupiter'' orbit has been investigated in Fogg and Nelson (2005,
2007) and Raymond et al. (2006b). These studies showed that a large
fraction of the disk's solid mass survives the inward migration of the
giant planet in two ways: (i) planetesimals are captured into mean
motion resonances interior to the orbit of the giant planet and, by
mutual collisions, give origin to massive terrestrial planets; (ii)
planetesimals are scattered into external orbits, where gas drag
re-circularizes their orbits; the standard terrestrial planet
formation process then resumes. Thus, the wide-spread expectation
that terrestrial planets could not exist in systems with a hot Jupiter
is not correct and future searches for extra-solar terrestrial planets
should not disregard these systems a priori. 

\subsection{Terrestrial planets evolution during giant planets instabilities}

The issue of terrestrial planet evolution during giant planets
instabilities deserves a whole section by itself. As usual, I start
with a description of our understanding of what happened in the solar
system.

As we have seen in Sect.~\ref{LHB}, we believe that the giant planets
of our solar system passed through a phase of orbital instability
$\sim 3.9$~Gy ago, i.e. well after the formation of the terrestrial
planets (which ended $\sim 4.5$~Gy ago; Kleine et al., 2009). During
this instability phase, close encounters between the giant planets
occurred; the orbits of the giant planets became eccentric and their
separation in semi major axis increased towards the current
values. During this period of chaotic evolution, a wide variety of
orbital histories are possible. We may, however, classify the orbital
histories in two classes: those in which Jupiter is not involved in
close encounters with another planet (only Saturn, Uranus and Neptune
have encounters with each other) and those in which Jupiter has
encounters with Uranus and/or Neptune (nicknamed below the {\it
jumping-Jupiter class}). The two classes give a very different
evolution of the orbital separation of Jupiter and Saturn.

In the first class of evolutions, the increase in orbital separation
between Jupiter and Saturn is due to planetesimal-driven migration.
In fact, if Saturn scatters an ice giant (Uranus or Neptune) while
Jupiter does not, necessarily Saturn has to scatter the ice giant
outwards and recoil towards the Sun. So, planetary encounters in this
class of evolutions lead to a reduction of the orbital separation of
Jupiter and Saturn and planetesimal-driven migration is the only
mechanism that can increase it. In the jumping-Jupiter class of
evolutions, most of the increase in orbital separation between the two
gas giants is instead due to planetary encounters. In fact, if Jupiter
has an encounter with an ice giant, said ice giant, given that it is
beyond the orbit of Saturn both at the beginning and at the end of the
evolution, must be first scattered inwards by Saturn and then be
scattered outwards by Jupiter. Thus, Saturn recoils outwards and
Jupiter inwards, which increases the orbital separation between
Jupiter and Saturn.

\begin{figure}[t!]
\centerline{\includegraphics[height=6.cm]{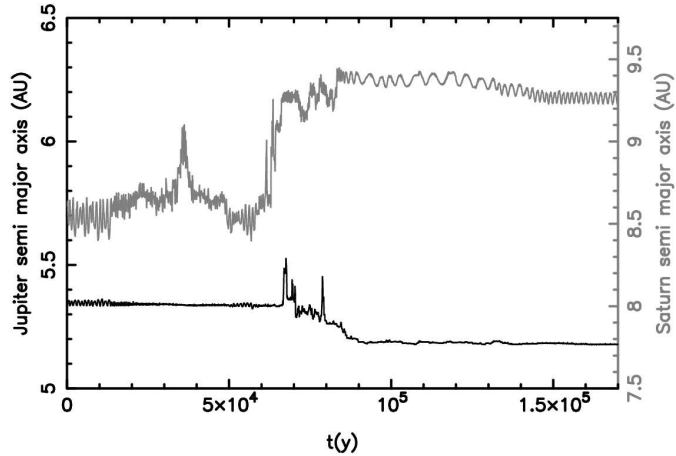}}
\vspace*{-.3cm} 
\caption{\small An example of ``jumping-Jupiter evolution''. The black
  and gray curves show the evolution of the semi major axes of Jupiter
  and Saturn, reported in the left-side and right-side vertical
  scales, respectively. The stochastic behavior is caused by
  encounters with a Uranus/Neptune-mass planet, originally placed the
  third in order of increasing distance from the Sun (not shown
  here). Time $t=0$ here is arbitrary and corresponds to the onset of
  the phase of planetary instability. The full evolution of the
  planets, which lasts 4.6 My, is illustrated in Fig.~4 of
  Brasser et al. (2009). All giant planets survived the full 4.6 My
  simulation on stable orbits, quite similar to those of the real
  planets of the Solar System.}
\label{JJ} 
\end{figure}

In summary, both classes of evolutions lead to an increase in the
orbital separation of Jupiter and Saturn, but the big difference is in
the timescale on which this separation occurs. Planetesimal-driven
migration is relatively slow: it forces the orbital separation to
evolve exponentially as $\Delta a(t)=\Delta a_{\rm
current}-\Delta_0\exp(-t/\tau)$, with $\tau\sim 5$--10~My (the
characteristic lifetime of planetesimals crossing the orbits of the
giant planets, such as the Centaur objects; Tiscareno and Malhotra,
2003; DiSisto and Brunini, 2007;
Bailey and Malhotra, 2009). Conversely, the phase of planetary
encounters is short, so that in the jumping-Jupiter class the orbital
separation of Jupiter and Saturn increases in less than $10^5$~y (see
Fig.~\ref{JJ}).

\begin{figure}[t!]
\centerline{\includegraphics[height=6.cm]{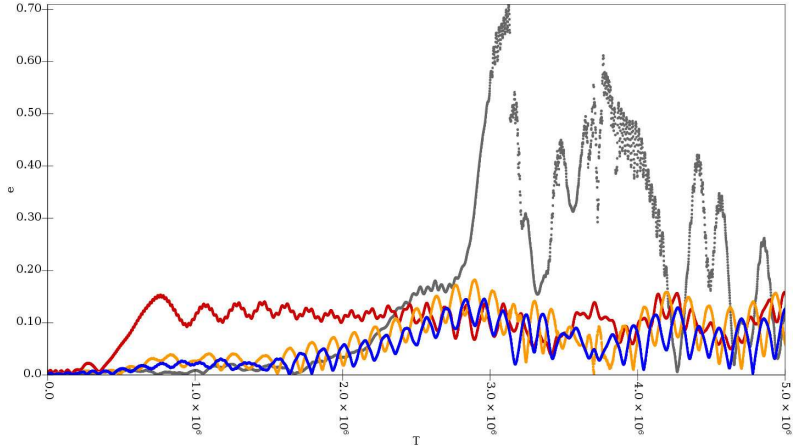}}
\vspace*{-.3cm} 
\caption{\small The evolutions of the eccentricities of Mars (red),
  Earth (blue), Venus (orange) and Mercury (grey), during
  planetesimal-drive migration of Jupiter and Saturn. Here the orbital
  separation between the two gas giants increases as $\Delta a_{\rm
current}-\Delta_0\exp(-t/\tau)$, with $\Delta_0=1.1$~AU and $\tau=5$~My.}
\label{Brasser} 
\end{figure}

The increase in orbital separation between Jupiter and Saturn changes
the secular frequencies of the orbits of these planets. If the
divergent migration of Jupiter and Saturn occurs on a timescale of a
few millions of years, as in the case of planetesimal-driven
migration, the orbits of the terrestrial planets are quite strongly
excited in eccentricity (Brasser et al., 2009). Even if starting from
circular orbits, the Earth and Venus acquire orbits whose
eccentricity oscillations exceed 0.15, i.e. twice as much as in the
real solar system; Mercury is destabilized (Fig.~\ref{Brasser}). This
happens because the frequency of precession of the perihelion of
Jupiter's orbit, denoted by $g_5$, decreases and, in sequence, it
becomes equal to those of Mars, Earth, Venus and Mercury
($g_4,\ldots,g_1$) respectively; every time that a frequency of a
terrestrial planet $g_k$ is equal to $g_5$ a secular resonance occurs
and the eccentricity of the corresponding planet is strongly
affected. Similarly, most of the asteroids in the inner part of the
asteroid belt are destabilized while the precession frequency of the
perihelion of the orbit of Saturn ($g_6$) decreases; consequently, the
final orbital distribution of the asteroids is incompatible with the
one currently observed (Morbidelli et al., 2010). In the jumping-Jupiter
class of evolutions, conversely, these problems do not exist because
the divergent migration of Jupiter and Saturn --and the consequent
decrease of $g_5$ and $g_6$-- are so fast that the eccentricities of
asteroids and terrestrial planets have no time to be seriously
affected. Thus, the terrestrial planets can have at the end orbital
eccentricities as small (or even smaller) than the current ones,
depending on their initial conditions (Brasser et al., 2009) and the
asteroid belt preserves roughly the orbital structure acquired during
terrestrial planets formation (Morbidelli et al., 2010).

The conclusion is that the real evolution of the giant planets of our
solar system had to be of the jumping-Jupiter class; otherwise the orbit of
the Earth, and the structure of the inner solar system in general,
would be substantially different from what they are now. 

There is no doubt that many evolutions of the giant planets can be
fatal for the formation or the evolution of habitable terrestrial
planets. Take the giant extra-solar planets discovered so far: most of
them have very eccentric orbits, which are thought to be the product
of a violent instability occurred in the original planetary system,
which led to close encounters between giant planets (see
Sect.~\ref{scatter}).  Raymond and Armitage (in preparation) show that
when giant planets acquire similarly large eccentricities, the
terrestrial planets in the system are forced to evolve onto orbits
with extreme eccentricities: many of them collide with the central
star, or start to intersect the orbits of the giant planets and are
then rapidly ejected onto hyperbolic orbits. Those terrestrial objects
which manage to survive, if any, do so on orbits with eccentricities that can
be hardly compatible with habitable worlds. Thus, the existence
of an habitable Earth in our system is possible only because our giant
planets remained on orbits with exceptionally small eccentricities
compared to the orbits of extra-solar planets. 

\section{Conclusions}

This chapter discussed the evolution of planetary systems.  The
emphasis has been put on the evolution of our solar system. Our team
effort, over the last 10 years, has been to reconstruct the history of
our system using computer simulations and taking advantage of all
possible detailed constraints (the orbits of the planets, the
architecture of the populations of small bodies, radioactive
chronologies for terrestrial planet formation, crater records etc.).
Although imperfect, I think that our view of the evolution of the
solar system, since the completion of giant planets formation, has
reached a quite satisfactory level of coherence. Conversely, the phase
of accretion of the giant planets remains poorly modeled. 

According to our understanding, the evolution of the solar system was
characterized by three main ``eras''. In the gas-disk era, Jupiter had
a wide-range radial migration. It first migrated inwards; then, when
it was at about 1.5~AU from the Sun, it got caught in resonance with
Saturn and, given the Jupiter/Saturn mass-ratio, it started to migrate
outwards (Walsh et al., 2011).  This inward-then-outward migration
explains why we do not have a ``hot (or warm) Jupiter'' in our solar
system. It also left indelible traces in the inner solar system,
particularly in the physical structure of the asteroid belt and the
small mass of Mars. As a result of the outward migration of Jupiter,
the four giant planets acquired a multi-resonant configuration, in
which each planet was in a mean-motion resonance with its neighbor.
The orbits of the giant planets were at the time much closer to each
other than they are now, and had significantly smaller eccentricities
and inclinations. 

At the disappearance of the gas, the system entered in the
planetesimal-disk era. The Earth and Venus completed their accretion
from a disk of planetary embryos and planetesimals that inherited an
outer edge at 1 AU from the earlier incursion of Jupiter into the
inner solar system. The accretion of Mars and of the asteroids was
frozen. Instead, a massive disk of planetesimals persisted outside the
orbit of the outermost giant planet. Its internal collisional activity
produced a debris disk comparable to those observed around $\sim 15$\%
of main sequence stars. Meanwhile, the gravitational interactions
between the giant planets and this disk, slowly modified the resonant
orbit of the former. Eventually, $\sim 600$~My later, the giant
planets became unstable, as a result of these slow orbital
modifications. The chaotic phase that followed reshuffled the
structure of the outer solar system: the giant planets acquired their
current orbits; most of the distant planetesimal disk was dispersed
and the Kuiper belt is what remains today of that disk (Levison et
al., 2008; Batygin et al., 2011).  Many asteroids also got released
from the asteroid belt. All these destabilized small bodies 
caused the Late Heavy Bombardment of the
terrestrial planets (Bottke et al., 2011). 

With this profound re-organization, the solar system entered into the
current era, lasting since $\sim 3.8$~Gy ago, in which it did not
suffer any further significant change.

If this story is true, then the evolution of our solar system was
defined by a sequence of specific features. For instance, the mass
ratio between Jupiter and Saturn prevented migration towards the Sun;
the late formation of the giant planets relative to the gas disk
lifetime prevented Saturn to grow more massive than Jupiter; the
instability phase that characterized the giant planets resulted in a
jumping-Jupiter evolution, which prevented secular resonances to
interfere with the orbits of the terrestrial planets, etcetera.  It
was then natural in this chapter to discuss what would have happened
if these events had not occurred, or if they had occurred
differently. This led me to address the origin of the 
diversity of planetary systems observed around other stars. The possible lines
of evolutions that I have described certainly do not exhaust all
possibilities: I'm sure that Nature has much more fantasy than we
have. However, they show that the evolution of a planetary system,
like the weather on Earth, is so sensitive on initial and
environmental conditions that a huge variety of outcomes is possible,
even starting from similar situations.

A frequently asked question, at this point, is whether one can predict
the probability that a system evolves towards one state or
another. The answer at this time, unfortunately, is no. Our
understanding of the first stages of planet formation is too limited;
moreover we don't know well enough the initial conditions, i.e. the
properties of proto-planetary disks. So, we cannot say which kinds of
planetary systems could be formed and with which probabilities;
without this information, we cannot estimate the probabilities of the
subsequent possible evolutions.  For instance, we don't know why
Jupiter and Saturn formed like they are, instead of having different
masses or having additional gas giant companions. In other words, we
can try to reconstruct the evolution of the solar system using all
available clues and constraints, but we are not able to ``predict'' our
solar system, a-priori. I'm not sure that we will ever be able to do
so. I think this field has a parallel with geology. Geologists are
able to reconstruct the complicated history of our continents with an
amazing precision, but they are not able to say what is the
probability that a terrestrial planet develops continents with the
properties of our own ones. We have to acknowledge that, at this time,
field of ``origins'' in planetary science is essentially a descriptive
discipline. As such, it is led by observations and interpreted by
theoretical models, not the other way around.

\acknowledgments

I am grateful to the Helmholtz Alliance's 'Planetary evolution and
life' and the French National Programme for Planetary Science, for
substantial financial support. I wish to thank all the people with
whom I worked over these fabulous years in the quest for a better
understanding of the evolution of planetary systems and of our solar
system in particular. I'm grateful to P. Michel, A. Crida, K. Walsh
and an anonymous reviewer, for carefully reading a first draft of this
chapter and for providing useful comments and suggestions.

\newpage
\centerline{\bf References}

\begin{itemize}

\item[$\bullet$] Abramov, O., 
Mojzsis, S.~J.\ 2009.\ Microbial habitability of the Hadean Earth during 
the late heavy bombardment.\ Nature 459, 419-422. 

\item[$\bullet$] Adams, F.~C., 
Laughlin, G.\ 2003.\ Migration and dynamical relaxation in crowded systems 
of giant planets.\ Icarus 163, 290-306. 

\item[$\bullet$] Agnor, C.~B., Canup, 
R.~M., Levison, H.~F.\ 1999.\ On the Character and Consequences of Large 
Impacts in the Late Stage of Terrestrial Planet Formation.\ Icarus 142, 
219-237. 

\item[$\bullet$] Agnor, C., Asphaug, 
E.\ 2004.\ Accretion Efficiency during Planetary Collisions.\ The 
Astrophysical Journal 613, L157-L160. 

\item[$\bullet$] Alibert, Y., Mordasini, C., Benz, W.\ 2004.\ Migration and giant planet formation.\ Astronomy and Astrophysics 417, L25-L28. 

\item[$\bullet$] Asphaug, E., Agnor, 
C.~B., Williams, Q.\ 2006.\ Hit-and-run planetary collisions.\ Nature 439, 
155-160. 

\item[$\bullet$] Bailey, B.~L., 
Malhotra, R.\ 2009.\ Two dynamical classes of Centaurs.\ Icarus 203, 
155-163. 

\item[$\bullet$] Baldwin, R.~B.\ 2006.\ Was 
there ever a Terminal Lunar Cataclysm?. With lunar viscosity arguments.\ 
Icarus 184, 308-318. 

\item[$\bullet$] Barge, P., Sommeria, J.\ 1995.\ Did planet formation begin inside persistent gaseous vortices?.\ Astronomy and Astrophysics 295, L1-L4.

\item[$\bullet$] Barnes, R., 
Greenberg, R.\ 2006.\ Stability Limits in Extrasolar Planetary Systems.\ 
The Astrophysical Journal 647, L163-L166. 

\item[$\bullet$] Baruteau, C., 
Masset, F.\ 2008.\ On the Corotation Torque in a Radiatively Inefficient 
Disk.\ The Astrophysical Journal 672, 1054-1067. 

\item[$\bullet$] Batygin, K., Brown, 
M.~E.\ 2010.\ Early Dynamical Evolution of the Solar System: Pinning Down 
the Initial Condition of the Nice Model.\ ArXiv e-prints
arXiv:1004.5414. 

\item[$\bullet$] Batygin, K., Brown, M.~E. and Fraser,
W.C. 2011. In-situ Formation of the Cold Classical Kuiper Belt. ApJ, sumbitted.

\item[$\bullet$] Bernstein, G.~M., 
Trilling, D.~E., Allen, R.~L., Brown, M.~E., Holman, M., Malhotra, R.\ 
2004.\ The Size Distribution of Trans-Neptunian Bodies.\ The Astronomical 
Journal 128, 1364-1390. 

\item[$\bullet$] Binzel, R.P., Bus, S.J., Burbine, T.H., Sunshine, J.M. 1996. Spectral Properties of Near-Earth Asteroids: Evidence for Sources of Ordinary Chondrite Meteorites. Science 273, 946-948. 

\item[$\bullet$] Bitsch, B., Kley, W.\ 2010.\ Orbital evolution of eccentric planets in
radiative discs.\ Astronomy and Astrophysics 523, A30.

\item[$\bullet$] Bodenheimer, P., 
Hubickyj, O., Lissauer, J.~J.\ 2000.\ Models of the in Situ Formation of 
Detected Extrasolar Giant Planets.\ Icarus 143, 2-14. 

\item[$\bullet$] Boley, A.~C.\ 2009.\ The Two 
Modes of Gas Giant Planet Formation.\ The Astrophysical Journal 695, 
L53-L57. 

\item[$\bullet$] Booth, M., Wyatt, M.~C., 
Morbidelli, A., Moro-Mart{\'{\i}}n, A., Levison, H.~F.\ 2009.\ The history 
of the Solar system's debris disc: observable properties of the Kuiper 
belt.\ Monthly Notices of the Royal Astronomical Society 399, 385-398. 

\item[$\bullet$] Boss, A.~P.\ 2000.\ Possible 
Rapid Gas Giant Planet Formation in the Solar Nebula and Other 
Protoplanetary Disks.\ The Astrophysical Journal 536, L101-L104. 

\item[$\bullet$] Boss, A.~P.\ 2001.\ Formation of 
Planetary-Mass Objects by Protostellar Collapse and Fragmentation.\ The 
Astrophysical Journal 551, L167-L170. 

\item[$\bullet$] Boss, A.~P.\ 2002.\ Stellar 
Metallicity and the Formation of Extrasolar Gas Giant Planets.\ The 
Astrophysical Journal 567, L149-L153. 

\item[$\bullet$] Bottke, W.~F., Levison, 
H.~F., Nesvorn{\'y}, D., Dones, L.\ 2007.\ Can planetesimals left over from 
terrestrial planet formation produce the lunar Late Heavy Bombardment?.\ 
Icarus 190, 203-223. 

\item[$\bullet$] Bottke, W.~F., Vokrouhlicky, D., Minton, D.,
  Nesvorny, D., Brasser, R., Simonson, B. 2011. The Great Archean
  Bombardment, or the Late Late Heavy Bombardment. Lunar and Planetary Institute Science Conference Abstracts 42, 2591.

\item[$\bullet$] Brasser, R., Morbidelli, A., Gomes, R., Tsiganis, K., Levison, H.~F.\ 2009.\ Constructing the secular architecture of the solar system II: the terrestrial planets.\ Astronomy and Astrophysics 507, 1053-1065.

\item[$\bullet$] Burbine, T.H., Binzel, R.P., Bus, S.J., Buchanan, P.C., Hinrichs, J.L., Hiroi, T., Meibom, A., Sunshine, J.M. 2000. Forging Asteroid-Meteorite Relationships Through Reflectance Spectroscopy. Lunar and Planetary Institute Science Conference Abstracts 31, 1844. 

\item[$\bullet$] Butler, R.~P., and 10 
colleagues 2006.\ Catalog of Nearby Exoplanets.\ The Astrophysical Journal 
646, 505-522. 

\item[$\bullet$] Cameron, A.~G.~W.\ 1978.\ Physics of the primitive
  solar accretion disk.\ Moon and Planets 18, 5-40. 

\item[$\bullet$] Canup, R.~M., 
Esposito, L.~W.\ 1996.\ Accretion of the Moon from an Impact-Generated 
Disk.\ Icarus 119, 427-446. 

\item[$\bullet$] Canup, R.~M., 
Asphaug, E.\ 2001.\ Origin of the Moon in a giant impact near the end of 
the Earth's formation.\ Nature 412, 708-712. 

\item[$\bullet$] Canup, R.~M., Ward, 
W.~R.\ 2006.\ A common mass scaling for satellite systems of gaseous 
planets.\ Nature 441, 834-839. 

\item[$\bullet$] Capobianco, C.~C., 
Duncan, M., Levison, H.~F.\ 2011.\ Planetesimal-driven planet migration in 
the presence of a gas disk.\ Icarus 211, 819-831. 

\item[$\bullet$] Carpenter, J.~M., and 
14 colleagues 2009.\ Formation and Evolution of Planetary Systems: 
Properties of Debris Dust Around Solar-Type Stars.\ The Astrophysical 
Journal Supplement Series 181, 197-226. 

\item[$\bullet$] Cassen, P.~M., Smith, 
B.~F., Miller, R.~H., Reynolds, R.~T.\ 1981.\ Numerical experiments on the 
stability of preplanetary disks.\ Icarus 48, 377-392. 

\item[$\bullet$] Chambers, 
J.~E., Wetherill, G.~W.\ 1998.\ Making the Terrestrial Planets: N-Body 
Integrations of Planetary Embryos in Three Dimensions.\ Icarus 136, 
304-327. 

\item[$\bullet$] Chambers, J.~E.\ 2001.\ 
Making More Terrestrial Planets.\ Icarus 152, 205-224. 

\item[$\bullet$] Chambers, J.~E., Cassen, P.\ 2002.\ The effects of nebula surface density profile and giant-planet 

\item[$\bullet$] Chambers, J.\ 2006.\ A 
semi-analytic model for oligarchic growth.\ Icarus 180, 496-513. 

\item[$\bullet$] Chatterjee, S., 
Ford, E.~B., Matsumura, S., Rasio, F.~A.\ 2008.\ Dynamical Outcomes of 
Planet-Planet Scattering.\ The Astrophysical Journal 686, 580-602. 

\item[$\bullet$] Chiang, E.~I.\ 2003.\ 
Excitation of Orbital Eccentricities by Repeated Resonance Crossings: 
Requirements.\ The Astrophysical Journal 584, 465-471. 

\item[$\bullet$] Cohen, B.~A., Swindle, 
T.~D., Kring, D.~A.\ 2000.\ Support for the Lunar Cataclysm Hypothesis from 
Lunar Meteorite Impact Melt Ages.\ Science 290, 1754-1756. 

\item[$\bullet$] Cresswell, P., Dirksen, G., Kley, W., Nelson, R.~P.\ 2007.\ On the evolution of eccentric and inclined protoplanets embedded in protoplanetary disks.\ Astronomy and Astrophysics 473, 329-342. 

\item[$\bullet$] Crida, A., Morbidelli, 
A., Masset, F.\ 2006.\ On the width and shape of gaps in protoplanetary 
disks.\ Icarus 181, 587-604. 

\item[$\bullet$] Crida, A., 
Morbidelli, A.\ 2007.\ Cavity opening by a giant planet in a protoplanetary 
disc and effects on planetary migration.\ Monthly Notices of the Royal 
Astronomical Society 377, 1324-1336. 

\item[$\bullet$] Crida, A., S{\'a}ndor, Z., Kley, W.\ 2008.\ Influence of an inner disc on the orbital evolution of massive planets migrating in resonance.\ Astronomy and Astrophysics 483, 325-337. 

\item[$\bullet$] Crida, A., Masset, F., 
Morbidelli, A.\ 2009.\ Long Range Outward Migration of Giant Planets, with 
Application to Fomalhaut b.\ The Astrophysical Journal 705, L148-L152. 

\item[$\bullet$] D'Angelo, G., Lubow, 
S.~H., Bate, M.~R.\ 2006.\ Evolution of Giant Planets in Eccentric Disks.\ 
The Astrophysical Journal 652, 1698-1714. 

\item[$\bullet$] Dauphas, N., 
Pourmand, A.\ 2011.\ Hf-W-Th evidence for rapid growth of Mars and its 
status as a planetary embryo.\ Nature 473, 489-492. 

\item[$\bullet$] Deutsch, A., 
Schaerer, U.\ 1994.\ Dating terrestrial impact events.\ Meteoritics 29, 
301-322. 

\item[$\bullet$] di Sisto, R.~P., 
Brunini, A.\ 2007.\ The origin and distribution of the Centaur population.\ 
Icarus 190, 224-235. 

\item[$\bullet$] Durisen, R.~H., Boss, 
A.~P., Mayer, L., Nelson, A.~F., Quinn, T., Rice, W.~K.~M.\ 2007.\ 
Gravitational Instabilities in Gaseous Protoplanetary Disks and 
Implications for Giant Planet Formation.\ Protostars and Planets V 607-622. 

\item[$\bullet$] Fernandez, J.~A., Ip, 
W.-H.\ 1984.\ Some dynamical aspects of the accretion of Uranus and Neptune 
- The exchange of orbital angular momentum with planetesimals.\ Icarus 58, 
109-120. 

\item[$\bullet$]  Ferraz-Mello, S., 
Beaug{\'e}, C., Michtchenko, T.~A.\ 2003.\ Evolution of Migrating Planet 
Pairs in Resonance.\ Celestial Mechanics and Dynamical Astronomy 87, 
99-112. 

\item[$\bullet$]  Fischer, D.~A., 
Valenti, J.\ 2005.\ The Planet-Metallicity Correlation.\ The Astrophysical 
Journal 622, 1102-1117. 

\item[$\bullet$] Fogg, M.~J., Nelson, R.~P.\ 2005.\ Oligarchic and giant impact growth of terrestrial planets in the presence of gas giant planet migration.\ Astronomy and Astrophysics 441, 791-806. 

\item[$\bullet$] Fogg, M.~J., Nelson, R.~P.\ 2007.\ On the formation of terrestrial planets in hot-Jupiter systems.\ Astronomy and Astrophysics 461, 1195-1208. 
\item[$\bullet$] Ford, E.~B., Havlickova, 
M., Rasio, F.~A.\ 2001.\ Dynamical Instabilities in Extrasolar Planetary 
Systems Containing Two Giant Planets.\ Icarus 150, 303-313. 

\item[$\bullet$] Ford, E.~B., Rasio, 
F.~A.\ 2008.\ Origins of Eccentric Extrasolar Planets: Testing the 
Planet-Planet Scattering Model.\ The Astrophysical Journal 686, 621-636. 

\item[$\bullet$] Fouchet, T., Moses, 
J.~I., Conrath, B.~J.\ 2009.\ Saturn: Composition and Chemistry.\ Saturn 
from Cassini-Huygens 83. 

\item[$\bullet$] Fuentes, C.~I., 
Holman, M.~J.\ 2008.\ a SUBARU Archival Search for Faint Trans-Neptunian 
Objects.\ The Astronomical Journal 136, 83-97. 

\item[$\bullet$] G{\'a}sp{\'a}r, 
A., Rieke, G.~H., Su, K.~Y.~L., Balog, Z., Trilling, D., Muzzerole, J., 
Apai, D., Kelly, B.~C.\ 2009.\ The Low Level of Debris Disk Activity at the 
Time of the Late Heavy Bombardment: A Spitzer Study of Praesepe.\ The 
Astrophysical Journal 697, 1578-1596. 

\item[$\bullet$] Goldreich, P., 
Tremaine, S.\ 1979.\ The excitation of density waves at the Lindblad and 
corotation resonances by an external potential.\ The Astrophysical Journal 
233, 857-871. 

\item[$\bullet$] Goldreich, P., 
Tremaine, S.\ 1980.\ Disk-satellite interactions.\ The Astrophysical 
Journal 241, 425-441. 

\item[$\bullet$] Goldreich, P., 
Sari, R.\ 2003.\ Eccentricity Evolution for Planets in Gaseous Disks.\ The 
Astrophysical Journal 585, 1024-1037. 

\item[$\bullet$] Goldreich, P., 
Lithwick, Y., Sari, R.\ 2004.\ Final Stages of Planet Formation.\ The 
Astrophysical Journal 614, 497-507. 


\item[$\bullet$] Gomes, R.~S., Morbidelli, 
A., Levison, H.~F.\ 2004.\ Planetary migration in a planetesimal disk: why 
did Neptune stop at 30 AU?.\ Icarus 170, 492-507. 

\item[$\bullet$] Gomes, R., Levison, 
H.~F., Tsiganis, K., Morbidelli, A.\ 2005.\ Origin of the cataclysmic Late 
Heavy Bombardment period of the terrestrial planets.\ Nature 435, 466-469. 

\item[$\bullet$] Gradie, J., Tedesco, E. 1982. Compositional structure of the
asteroid belt. Science 216, 1405-1407. 

\item[$\bullet$] Greenberg, R., 
Hartmann, W.~K., Chapman, C.~R., Wacker, J.~F.\ 1978.\ Planetesimals to 
planets - Numerical simulation of collisional evolution.\ Icarus 35, 1-26. 

\item[$\bullet$] Greenzweig, 
Y., Lissauer, J.~J.\ 1992.\ Accretion rates of protoplanets. II - Gaussian 
distributions of planetesimal velocities.\ Icarus 100, 440-463. 

\item[$\bullet$] Guillot, T.\ 2005.\ THE 
INTERIORS OF GIANT PLANETS: Models and Outstanding Questions.\ Annual 
Review of Earth and Planetary Sciences 33, 493-530. 

\item[$\bullet$] Guillot, T., Hueso, 
R.\ 2006.\ The composition of Jupiter: sign of a (relatively) late 
formation in a chemically evolved protosolar disc.\ Monthly Notices of the 
Royal Astronomical Society 367, L47-L51. 

\item[$\bullet$] Guillot, T., Santos, N.~C., Pont, F., Iro, N., Melo, C., Ribas, I.\ 2006.\ A correlation between the heavy element content of transiting extrasolar planets and the metallicity of their parent stars.\ Astronomy and Astrophysics 453, L21-L24. 

\item[$\bullet$] Haisch, K.~E., Jr., 
Lada, E.~A., Lada, C.~J.\ 2001.\ Disk Frequencies and Lifetimes in Young 
Clusters.\ The Astrophysical Journal 553, L153-L156. 

\item[$\bullet$] Hansen, B.~M.~S.\ 2009.\ 
Formation of the Terrestrial Planets from a Narrow Annulus.\ The 
Astrophysical Journal 703, 1131-1140. 

\item[$\bullet$] Hartmann, W.~K., 
Ryder, G., Dones, L., Grinspoon, D.\ 2000.\ The Time-Dependent Intense 
Bombardment of the Primordial Earth/Moon System.\ Origin of the earth and 
moon, edited by R.M.~Canup and K.~Righter and 69 collaborating 
authors.~Tucson: University of Arizona Press., p.493-512 493-512. 

\item[$\bullet$] Hartmann, W.~K., 
Quantin, C., Mangold, N.\ 2007.\ Possible long-term decline in impact 
rates. 2. Lunar impact-melt data regarding impact history.\ Icarus 186, 
11-23. 

\item[$\bullet$] Hayashi, C.\ 1981. Structure of the Solar
Nebula, Growth and Decay of Magnetic Fields and Effects of Magnetic
and Turbulent Viscosities on the Nebula.\ {\it Progress of Theoretical
Physics Supplement} 70, 35-53.

\item[$\bullet$] Henrard, J. 1993. The adiabatic invariants in classical
mechanics, {\it Dynamics Reported} {2}, 117--235.

\item[$\bullet$] Hillenbrand, L.~A., 
Carpenter, J.~M., Kim, J.~S., Meyer, M.~R., Backman, D.~E., 
Moro-Mart{\'{\i}}n, A., Hollenbach, D.~J., Hines, D.~C., Pascucci, I., 
Bouwman, J.\ 2008.\ The Complete Census of 70 {$\mu$}m-bright Debris Disks 
within ``the Formation and Evolution of Planetary Systems'' Spitzer Legacy 
Survey of Sun-like Stars.\ The Astrophysical Journal 677, 630-656. 

\item[$\bullet$] Ida, S., Makino, J.\ 
1993.\ Scattering of planetesimals by a protoplanet - Slowing down of 
runaway growth.\ Icarus 106, 210. 

\item[$\bullet$] Ida, S., Bryden, G., Lin, 
D.~N.~C., Tanaka, H.\ 2000.\ Orbital Migration of Neptune and Orbital 
Distribution of Trans-Neptunian Objects.\ The Astrophysical Journal 534, 
428-445. 

\item[$\bullet$] Ida, S., Lin, D.~N.~C.\ 
2004.\ Toward a Deterministic Model of Planetary Formation. II. The 
Formation and Retention of Gas Giant Planets around Stars with a Range of 
Metallicities.\ The Astrophysical Journal 616, 567-572. 

\item[$\bullet$] Ida, S., Lin, D.~N.~C.\ 
2008.\ Toward a Deterministic Model of Planetary Formation. V. Accumulation 
Near the Ice Line and Super-Earths.\ The Astrophysical Journal 685, 
584-595. 

\item[$\bullet$] Juri{\'c}, M., 
Tremaine, S.\ 2008.\ Dynamical Origin of Extrasolar Planet Eccentricity 
Distribution.\ The Astrophysical Journal 686, 603-620. 

\item[$\bullet$] Kalas, P., Graham, J.~R., 
Chiang, E., Fitzgerald, M.~P., Clampin, M., Kite, E.~S., Stapelfeldt, K., 
Marois, C., Krist, J.\ 2008.\ Optical Images of an Exosolar Planet 25 
Light-Years from Earth.\ Science 322, 1345. 

\item[$\bullet$] Kelsall, T., and 11 
colleagues 1998.\ The COBE Diffuse Infrared Background Experiment Search 
for the Cosmic Infrared Background. II. Model of the Interplanetary Dust 
Cloud.\ The Astrophysical Journal 508, 44-73. 

\item[$\bullet$]  Kenyon, S.~J., 
Bromley, B.~C.\ 2006.\ Terrestrial Planet Formation. I. The Transition from 
Oligarchic Growth to Chaotic Growth.\ The Astronomical Journal 131, 
1837-1850. 

\item[$\bullet$] Kenyon, S.~J., Bromley, 
B.~C., O'Brien, D.~P., Davis, D.~R.\ 2008.\ Formation and Collisional 
Evolution of Kuiper Belt Objects.\ The Solar System Beyond Neptune 293-313. 

\item[$\bullet$] Kirsh, D.~R., Duncan, M., 
Brasser, R., Levison, H.~F.\ 2009.\ Simulations of planet migration driven 
by planetesimal scattering.\ Icarus 199, 197-209. 

\item[$\bullet$] Kleine, T., Touboul, M., 
Bourdon, B., Nimmo, F., Mezger, K., Palme, H., Jacobsen, S.~B., Yin, Q.-Z., 
Halliday, A.~N.\ 2009.\ Hf-W chronology of the accretion and early 
evolution of asteroids and terrestrial planets.\ Geochimica et Cosmochimica 
Acta 73, 5150-5188. 

\item[$\bullet$] Kley, W., Peitz, J., Bryden, G.\ 2004.\ Evolution of planetary systems in resonance.\ Astronomy and Astrophysics 414, 735-747. 

\item[$\bullet$] Kley, W., Lee, M.~H., Murray, N., Peale, S.~J.\ 2005.\ Modeling the resonant planetary system GJ 876.\ Astronomy and Astrophysics 437, 727-742. 
\item[$\bullet$]Kley, W., Dirksen, G.\ 2006.\ Disk eccentricity and embedded planets.\ Astronomy and Astrophysics 447, 369-377. 

\item[$\bullet$] Kley, W., Crida, A.\ 2008.\ Migration of protoplanets in radiative discs.\ Astronomy and Astrophysics 487, L9-L12. 

\item[$\bullet$] Koeberl, C.\ 2004. The late heavy bombardment in the
  inner solar system: is there any connection to Kuiper belt objects?
  The first decadal review of the Edgeworth-Kuiper Belt. 79--87.

\item[$\bullet$] Kokubo, E., Ida, S.\ 
1998.\ Oligarchic Growth of Protoplanets.\ Icarus 131, 171-178. 

\item[$\bullet$] Kokubo, E., Kominami, 
J., Ida, S.\ 2006.\ Formation of Terrestrial Planets from Protoplanets. I. 
Statistics of Basic Dynamical Properties.\ The Astrophysical Journal 642, 
1131-1139. 

\item[$\bullet$] Kokubo, E., Genda, 
H.\ 2010.\ Formation of Terrestrial Planets from Protoplanets Under a 
Realistic Accretion Condition.\ The Astrophysical Journal 714, L21-L25. 

\item[$\bullet$] Kring, D.~A., Cohen, 
B.~A.\ 2002.\ Cataclysmic bombardment throughout the inner solar system 
3.9-4.0 Ga.\ Journal of Geophysical Research (Planets) 107, 5009. 

\item[$\bullet$] Levison, H.~F., 
Lissauer, J.~J., Duncan, M.~J.\ 1998.\ Modeling the Diversity of Outer 
Planetary Systems.\ The Astronomical Journal 116, 1998-2014. 


\item[$\bullet$] Levison, H.~F., Dones, 
L., Chapman, C.~R., Stern, S.~A., Duncan, M.~J., Zahnle, K.\ 2001.\ Could 
the Lunar ``Late Heavy Bombardment'' Have Been Triggered by the Formation 
of Uranus and Neptune?.\ Icarus 151, 286-306. 

\item[$\bullet$] Levison, H.~F., 
Agnor, C.\ 2003.\ The Role of Giant Planets in Terrestrial Planet 
Formation.\ The Astronomical Journal 125, 2692-2713. 

\item[$\bullet$] Levison, H.~F., 
Morbidelli, A.\ 2007.\ Models of the collisional damping scenario for 
ice-giant planets and Kuiper belt formation.\ Icarus 189, 196-212. 

\item[$\bullet$] Levison, H.~F., 
Morbidelli, A., Gomes, R., Backman, D.\ 2007.\ Planet Migration in 
Planetesimal Disks.\ Protostars and Planets V 669-684. 

\item[$\bullet$] Levison, H.~F., 
Morbidelli, A., Vanlaerhoven, C., Gomes, R., Tsiganis, K.\ 2008.\ Origin of 
the structure of the Kuiper belt during a dynamical instability in the 
orbits of Uranus and Neptune.\ Icarus 196, 258-273. 

\item[$\bullet$] Levison, H.~F., Bottke, 
W.~F., Gounelle, M., Morbidelli, A., Nesvorn{\'y}, D., Tsiganis, K.\ 2009.\ 
Contamination of the asteroid belt by primordial trans-Neptunian objects.\ 
Nature 460, 364-366. 

\item[$\bullet$] Levison, H.~F., 
Thommes, E., Duncan, M.~J.\ 2010.\ Modeling the Formation of Giant Planet 
Cores. I. Evaluating Key Processes.\ The Astronomical Journal 139, 
1297-1314. 

\item[$\bullet$] Levison, H.~F., Morbidelli, A., Tsiganis, K., Nesvorny,
  D. and Gomes, R. 2011.  Late orbital instabilities in
  the outer planets induced by interaction with a self-gravitating
  planetesimal disk. To be submitted.

\item[$\bullet$] Lin, D.~N.~C., 
Papaloizou, J.\ 1986a.\ On the tidal interaction between protoplanets and 
the primordial solar nebula. II - Self-consistent nonlinear interaction.\ 
The Astrophysical Journal 307, 395-409. 

\item[$\bullet$] Lin, D.~N.~C., 
Papaloizou, J.\ 1986.\ On the tidal interaction between protoplanets and 
the protoplanetary disk. III - Orbital migration of protoplanets.\ The 
Astrophysical Journal 309, 846-857. 

\item[$\bullet$] Lin, D.~N.~C., Bodenheimer, 
P., Richardson, D.~C.\ 1996.\ Orbital migration of the planetary companion 
of 51 Pegasi to its present location.\ Nature 380, 606-607. 

\item[$\bullet$] Lin, D.~N.~C., Ida, S.\ 
1997.\ On the Origin of Massive Eccentric Planets.\ The Astrophysical 
Journal 477, 781. 

\item[$\bullet$] Lodders, K.\ 2003.\ Solar 
System Abundances and Condensation Temperatures of the Elements.\ The 
Astrophysical Journal 591, 1220-1247. 

\item[$\bullet$] Lynden-Bell, 
D., Pringle, J.~E.\ 1974.\ The evolution of viscous discs and the origin of 
the nebular variables..\ Monthly Notices of the Royal Astronomical Society 
168, 603-637. 

\item[$\bullet$] Lyra, W., Johansen, A., Klahr, H., Piskunov, N.\
  2009a.\ Standing on the shoulders of giants. Trojan Earths and
  vortex trapping in low mass self-gravitating protoplanetary disks of
  gas and solids.\ Astronomy and Astrophysics 493, 1125-1139.

\item[$\bullet$] Lyra, W., Johansen, A., Zsom, A., Klahr, H.,
Piskunov, N.\ 2009b.\ Planet formation bursts at the borders of the
dead zone in 2D numerical simulations of circumstellar disks.\
Astronomy and Astrophysics 497, 869-888.

\item[$\bullet$] Lyra, W., Paardekooper, 
S.-J., Mac Low, M.-M.\ 2010.\ Orbital Migration of Low-mass Planets in 
Evolutionary Radiative Models: Avoiding Catastrophic Infall.\ The 
Astrophysical Journal 715, L68-L73. 

\item[$\bullet$] Malhotra, R.\ 1993.\ The 
origin of Pluto's peculiar orbit.\ Nature 365, 819-821. 

\item[$\bullet$] Malhotra, R.\ 1995.\ The 
Origin of Pluto's Orbit: Implications for the Solar System Beyond Neptune.\ 
The Astronomical Journal 110, 420. 

\item[$\bullet$] Marois, C., Macintosh, 
B., Barman, T., Zuckerman, B., Song, I., Patience, J., Lafreni{\`e}re, D., 
Doyon, R.\ 2008.\ Direct Imaging of Multiple Planets Orbiting the Star HR 
8799.\ Science 322, 1348. 

\item[$\bullet$] Marzari, 
F., Weidenschilling, S.~J.\ 2002.\ Eccentric Extrasolar Planets: The 
Jumping Jupiter Model.\ Icarus 156, 570-579. 

\item[$\bullet$] Marzari, F., Baruteau, C., Scholl, H.\ 2010.\ Planet-planet scattering
in circumstellar gas disks.\ Astronomy and Astrophysics 514, L4. 

\item[$\bullet$] Masset, F., 
Snellgrove, M.\ 2001.\ Reversing type II migration: resonance trapping of a 
lighter giant protoplanet.\ Monthly Notices of the Royal Astronomical 
Society 320, L55-L59. 

\item[$\bullet$] Masset, F.~S., 
Morbidelli, A., Crida, A., Ferreira, J.\ 2006.\ Disk Surface Density 
Transitions as Protoplanet Traps.\ The Astrophysical Journal 642, 478-487. 

\item[$\bullet$] Milani, A., Nobili, A.~M., Carpino, M.\ 1987.\ Secular variations of the semimajor axes - Theory and experiments.\ Astronomy and Astrophysics 172, 265-279. 

\item[$\bullet$] Militzer, B., Hubbard, W.~B.\ 2009.\ Comparison of Jupiter interior models derived from first-principles simulations.\ Astrophysics and Space Science 322, 129-133. 

\item[$\bullet$] Min, M., Dullemond, C.~P., 
Kama, M., Dominik, C.\ 2011.\ The thermal structure and the location of the 
snow line in the protosolar nebula: Axisymmetric models with full 3-D 
radiative transfer.\ Icarus 212, 416-426. 

\item[$\bullet$] Minton, D.~A., 
Malhotra, R.\ 2009.\ A record of planet migration in the main asteroid 
belt.\ Nature 457, 1109-1111. 

\item[$\bullet$] Moekel, N., Raymond, S.N., Armitage,
  Ph.J. 2008. Extrasolar planets eccentricities from scattering in the
  presence of residual gas-disks. Ap.J. 688, 1361-1367

\item[$\bullet$] Moorhead, A.~V., 
Adams, F.~C.\ 2005.\ Giant planet migration through the action of disk 
torques and planet planet scattering.\ Icarus 178, 517-539. 

\item[$\bullet$] Morbidelli, A., Chambers, J., Lunine, J.~I., Petit, J.~M., Robert, F.,
Valsecchi, G.~B., Cyr, K.~E.\ 2000.\ Source regions and time scales for the delivery of water to Earth.\ Meteoritics and Planetary Science 35, 1309-1320. 

\item[$\bullet$] Morbidelli, A., 
Crida, A.\ 2007.\ The dynamics of Jupiter and Saturn in the gaseous 
protoplanetary disk.\ Icarus 191, 158-171. 

\item[$\bullet$] Morbidelli, A., 
Tsiganis, K., Crida, A., Levison, H.~F., Gomes, R.\ 2007.\ Dynamics of the 
Giant Planets of the Solar System in the Gaseous Protoplanetary Disk and 
Their Relationship to the Current Orbital Architecture.\ The Astronomical 
Journal 134, 1790-1798. 

\item[$\bullet$] Morbidelli, A., Crida, A., Masset, F., Nelson, R.~P.\
  2008.\ Building giant-planet cores at a planet trap.\ Astronomy and
  Astrophysics 478, 929-937. 

\item[$\bullet$] Morbidelli, A., 
Levison, H.~F., Gomes, R.\ 2008b.\ The Dynamical Structure of the Kuiper 
Belt and Its Primordial Origin.\ The Solar System Beyond Neptune 275-292.

\item[$\bullet$] Morbidelli, A., Brasser, R., Gomes, R., Levison
  H.F. and Tsiganis, K., 2010. Evidence from the asteroid belt for a
  violent past evolution of Jupiter's orbit. Astron. J., submitted.  

\item[$\bullet$] Morishima, R., 
Stadel, J., Moore, B.\ 2010.\ From planetesimals to terrestrial planets: 
N-body simulations including the effects of nebular gas and giant planets.\ 
Icarus 207, 517-535. 

\item[$\bullet$] Morfill, G.~E., 
Voelk, H.~J.\ 1984.\ Transport of dust and vapor and chemical fractionation 
in the early protosolar cloud.\ The Astrophysical Journal 287, 371-395. 

\item[$\bullet$] Nelson, R.~P., 
Papaloizou, J.~C.~B.\ 2003.\ The interaction of a giant planet with a disc 
with MHD turbulence - II. The interaction of the planet with the disc.\ 
Monthly Notices of the Royal Astronomical Society 339, 993-1005. 

\item[$\bullet$] Nelson, R.~P.\ 2005.\ On the orbital evolution of low mass protoplanets in turbulent, magnetised disks.\ Astronomy and Astrophysics 443, 1067-1085. 

\item[$\bullet$] Nesvorn{\'y}, D., 
Jenniskens, P., Levison, H.~F., Bottke, W.~F., Vokrouhlick{\'y}, D., 
Gounelle, M.\ 2010.\ Cometary Origin of the Zodiacal Cloud and Carbonaceous 
Micrometeorites. Implications for Hot Debris Disks.\ The Astrophysical 
Journal 713, 816-836. 

\item[$\bullet$] Nettelmann, N., 
Holst, B., Kietzmann, A., French, M., Redmer, R., Blaschke, D.\ 2008.\ Ab 
Initio Equation of State Data for Hydrogen, Helium, and Water and the 
Internal Structure of Jupiter.\ The Astrophysical Journal 683, 1217-1228. 

\item[$\bullet$] O'Brien, D.~P., 
Morbidelli, A., Levison, H.~F.\ 2006.\ Terrestrial planet formation with 
strong dynamical friction.\ Icarus 184, 39-58. 

\item[$\bullet$] O'Brien, D.~P., 
Morbidelli, A., Bottke, W.~F.\ 2007.\ The primordial excitation and 
clearing of the asteroid belt - Revisited.\ Icarus 191, 434-452. 

\item[$\bullet$] \"Opik, E. J., 1976.  {\it Interplanetary Encounters:
Close Range
Gravitational Interactions}.  Elsevier, New York.

\item[$\bullet$] Paardekooper, S.-J., Mellema, G.\ 2006.\ Halting type I planet migration in non-isothermal disks.\ Astronomy and Astrophysics 459, L17-L20. 

\item[$\bullet$] Paardekooper, S.-J., Papaloizou, J.~C.~B.\ 2009.\ On corotation torques, 
horseshoe drag and the possibility of sustained stalled or outward 
protoplanetary migration.\ Monthly Notices of the Royal Astronomical 
Society 394, 2283-2296. 

\item[$\bullet$] Paardekooper, 
S.-J., Baruteau, C., Crida, A., Kley, W.\ 2010.\ A torque formula for 
non-isothermal type I planetary migration - I. Unsaturated horseshoe drag.\ 
Monthly Notices of the Royal Astronomical Society 401, 1950-1964. 

\item[$\bullet$] Papaloizou, J.~C.~B., Nelson, R.~P., Masset, F.\ 2001.\ Orbital eccentricity growth through disc-companion tidal interaction.\ Astronomy and Astrophysics 366, 263-275.

\item[$\bullet$] Petit, J.-M., Morbidelli, 
A., Chambers, J.\ 2001.\ The Primordial Excitation and Clearing of the 
Asteroid Belt.\ Icarus 153, 338-347. 

\item[$\bullet$] Pierens, A., Nelson, R.~P.\ 2008.\ Constraints on resonant-trapping for two planets embedded in a protoplanetary disc.\ Astronomy and Astrophysics 482, 333-340.

\item[$\bullet$] Podolak, M., Zucker, S.\ 2004.\ A note on the snow line in protostellar accretion disks.\ Meteoritics and Planetary Science 39, 1859-1868. 

\item[$\bullet$] Pollack, J.~B., 
Hubickyj, O., Bodenheimer, P., Lissauer, J.~J., Podolak, M., Greenzweig, 
Y.\ 1996.\ Formation of the Giant Planets by Concurrent Accretion of Solids 
and Gas.\ Icarus 124, 62-85. 

\item[$\bullet$] Rafikov, R.~R.\ 2004.\ Fast 
Accretion of Small Planetesimals by Protoplanetary Cores.\ The Astronomical 
Journal 128, 1348-1363. 

\item[$\bullet$] Rasio, F.~A., Ford, 
E.~B.\ 1996.\ Dynamical instabilities and the formation of extrasolar 
planetary systems.\ Science 274, 954-956. 

\item[$\bullet$] Raymond, S.~N., Quinn, 
T., Lunine, J.~I.\ 2004.\ Making other earths: dynamical simulations of 
terrestrial planet formation and water delivery.\ Icarus 168, 1-17. 

\item[$\bullet$] Raymond, S.~N., Quinn, 
T., Lunine, J.~I.\ 2005.\ Terrestrial Planet Formation in Disks with 
Varying Surface Density Profiles.\ The Astrophysical Journal 632,
670-676. 

\item[$\bullet$] Raymond, S.~N., Quinn, 
T., Lunine, J.~I.\ 2006.\ High-resolution simulations of the final assembly 
of Earth-like planets I. Terrestrial accretion and dynamics.\ Icarus 183, 
265-282. 

\item[$\bullet$] Raymond, S.~N., 
Mandell, A.~M., Sigurdsson, S.\ 2006b.\ Exotic Earths: Forming Habitable 
Worlds with Giant Planet Migration.\ Science 313, 1413-1416. 

\item[$\bullet$] Raymond, S.~N., Quinn, 
T., Lunine, J.~I.\ 2007.\ High-Resolution Simulations of The Final Assembly 
of Earth-Like Planets. 2. Water Delivery And Planetary Habitability.\ 
Astrobiology 7, 66-84. 

\item[$\bullet$] Raymond, S.~N., Barnes, 
R., Veras, D., Armitage, P.~J., Gorelick, N., Greenberg, R.\ 2009.\ 
Planet-Planet Scattering Leads to Tightly Packed Planetary Systems.\ The 
Astrophysical Journal 696, L98-L101. 

\item[$\bullet$] Raymond, S.~N., 
O'Brien, D.~P., Morbidelli, A., Kaib, N.~A.\ 2009b.\ Building the 
terrestrial planets: Constrained accretion in the inner Solar System.\ 
Icarus 203, 644-662. 

\item[$\bullet$] Ryder, G., 1990, Lunar samples, lunar accretion and the Early bombardment of
the Moon: Eos Transactions AGU, v. 71, p. 313-323.

\item[$\bullet$] Ryder, G., Koeberl, C., and Mojzsis, S.J., 2000, Heavy bombardment on the Earth at ~ 3.85: The search for petrographical and geochemical evidence, in Canup, R.M., and Righter, K., eds., Origin of the Earth and Moon: Tucson, Arizona, University of Arizona Press, p. 475-492.

\item[$\bullet$] Ryder, G., 2002, Mass flux in the ancient Earth-Moon system and the benign
implications for the origin of life on Earth: Journal Geophysical
Research-Planets, v. 107, p. 6-14.

\item[$\bullet$] S{\'a}ndor, Z., 
Lyra, W., Dullemond, C.~P.\ 2011.\ Formation of Planetary Cores at Type I 
Migration Traps.\ The Astrophysical Journal 728, L9. 

\item[$\bullet$] Saslaw, W.~C.\ 1985.\ 
Thermodynamics and galaxy clustering - Relaxation of N-body experiments.\ 
The Astrophysical Journal 297, 49-60. 

\item[$\bullet$] Stamatellos, D., Whitworth, A.~P.\ 2008.\ Can giant
  planets form by gravitational fragmentation of discs? \ Astronomy and Astrophysics 480, 879-887. 

\item[$\bullet$] Stewart, G., 
Wetherill, G.\ 1988.\ Evolution of planetesimal velocities.\ Icarus 79, 
542-553. 

\item[$\bullet$] Shakura, N.I. and Sunyaev, R.A. 1973. Black holes in
  binary systems. Observational appearance. Astronomy and Astrophysics 24, 337--355.

\item[$\bullet$] Tanaka, H., Takeuchi, 
T., Ward, W.~R.\ 2002.\ Three-Dimensional Interaction between a Planet and 
an Isothermal Gaseous Disk. I. Corotation and Lindblad Torques and Planet 
Migration.\ The Astrophysical Journal 565, 1257-1274. 

\item[$\bullet$] Tera, F., Papanastassiou, D.~A., Wasserburg, G.~J.\
  1974.\ Isotopic evidence for a terminal lunar cataclysm.\ Earth and
  Planetary Science Letters 22, 1.

\item[$\bullet$] Thommes, E.~W., Duncan, 
M.~J., Levison, H.~F.\ 1999.\ The formation of Uranus and Neptune in the 
Jupiter-Saturn region of the Solar System.\ Nature 402, 635-638. 

\item[$\bullet$] Thommes, E.~W., Duncan, 
M.~J., Levison, H.~F.\ 2003.\ Oligarchic growth of giant planets.\ Icarus 
161, 431-455. 

\item[$\bullet$] Thommes, E., Nagasawa, 
M., Lin, D.~N.~C.\ 2008.\ Dynamical Shake-up of Planetary Systems. II. 
N-Body Simulations of Solar System Terrestrial Planet Formation Induced by 
Secular Resonance Sweeping.\ The Astrophysical Journal 676, 728-739. 

\item[$\bullet$] Tiscareno, 
M.~S., Malhotra, R.\ 2003.\ The Dynamics of Known Centaurs.\ The 
Astronomical Journal 126, 3122-3131. 

\item[$\bullet$] Trail, D., Mojzsis, 
S.~J., Harrison, T.~M.\ 2007.\ Thermal events documented in Hadean zircons 
by ion microprobe depth profiles.\ Geochimica et Cosmochimica Acta 71, 
4044-4065. 

\item[$\bullet$] Tsiganis, K., Gomes, 
R., Morbidelli, A., Levison, H.~F.\ 2005.\ Origin of the orbital 
architecture of the giant planets of the Solar System.\ Nature 435, 
459-461. 

\item[$\bullet$] Valley J. W., Peck W. H., King E. M., and Wilde S. A.
   2002. A cool early Earth. Geology 30, 351-354.

\item[$\bullet$] Valsecchi, A., Manara, G.~B.\ 1997.\ Dynamics of comets in the outer planetary region. II. Enhanced planetary masses and orbital evolutionary
 paths.\ Astronomy and Astrophysics 323, 986-998.

\item[$\bullet$] Veras, D., 
Armitage, P.~J.\ 2004.\ Outward migration of extrasolar planets to large 
orbital radii.\ Monthly Notices of the Royal Astronomical Society 347, 
613-624. 

\item[$\bullet$] Veras, D., Crepp, J.~R., 
Ford, E.~B.\ 2009.\ Formation, Survival, and Detectability of Planets 
Beyond 100 AU.\ The Astrophysical Journal 696, 1600-1611. 

\item[$\bullet$] Walsh, K.J., Morbidelli, A., Raymond, S.N. O'Brien,
  D.P., Mandell, A.M. 2011. A low mass for Mars from Jupiter's early
  gas-driven migration. Nature, in press.

\item[$\bullet$] Ward, W.~R.\ 1986.\ Density waves 
in the solar nebula - Differential Lindblad torque.\ Icarus 67, 164-180. 

\item[$\bullet$] Ward, W.~R.\ 1997.\ Protoplanet 
Migration by Nebula Tides.\ Icarus 126, 261-281. 

\item[$\bullet$] Weidenschilling, S.~J.\ 1977.\ The distribution of mass in the planetary system and solar nebula.\ Astrophysics and Space Science 51, 153-158. 
\item[$\bullet$] Weidenschilling, S.~J., Marzari, F.\ 1996.\ Gravitational scattering as a 
possible origin for giant planets at small stellar distances.\ Nature 384, 
619-621. 

\item[$\bullet$] Weidenschilling, S.~J., Spaute, D., Davis, D.~R., Marzari, F., Ohtsuki, K.\ 
1997.\ Accretional Evolution of a Planetesimal Swarm.\ Icarus 128, 429-455. 

\item[$\bullet$] Wetherill, 
G.~W., Stewart, G.~R.\ 1989.\ Accumulation of a swarm of small 
planetesimals.\ Icarus 77, 330-357. 

\item[$\bullet$] Wetherill, G.~W.\ 1992.\ An 
alternative model for the formation of the asteroids.\ Icarus 100, 307-325. 

\item[$\bullet$] Wetherill, 
G.~W., Stewart, G.~R.\ 1993.\ Formation of planetary embryos - Effects of 
fragmentation, low relative velocity, and independent variation of 
eccentricity and inclination.\ Icarus 106, 190. 

\item[$\bullet$] Villeneuve, J., Chaussidon, M., Libourel,
G. 2009. Homogeneous Distribution of 26Al in the Solar System from the
Mg Isotopic Composition of Chondrules. Science 325, 985-988.

\item[$\bullet$] Wong, M.~H., Mahaffy, 
P.~R., Atreya, S.~K., Niemann, H.~B., Owen, T.~C.\ 2004.\ Updated Galileo 
probe mass spectrometer measurements of carbon, oxygen, nitrogen, and 
sulfur on Jupiter.\ Icarus 171, 153-170. 

\item[$\bullet$] Wyatt, M.~C., Smith, R., 
Greaves, J.~S., Beichman, C.~A., Bryden, G., Lisse, C.~M.\ 2007.\ 
Transience of Hot Dust around Sun-like Stars.\ The Astrophysical Journal 
658, 569-583. 

\item[$\bullet$] Zhang, H., Zhou, J.-L.\ 
2010.\ On the Orbital Evolution of a Giant Planet Pair Embedded in a 
Gaseous Disk. I. Jupiter-Saturn Configuration.\ The Astrophysical Journal 
714, 532-548. 

\end{itemize}

\end{document}